\documentclass[fleqn,usenatbib]{mnras}

\usepackage[T1]{fontenc}
\usepackage{ae,aecompl}

\usepackage{graphicx}	
\usepackage{amsmath}	
\usepackage{amssymb}	

\usepackage{xcolor}

\hypersetup{draft}

\title{Deep Long Asymmetric Occultation in EPIC 204376071}

\author[Rappaport et al.]{
S.~Rappaport$^1$,   
G.~Zhou$^{2,3}$,
A.~Vanderburg$^{4,5}$, 
A.~Mann$^{6}$,
M.H.~Kristiansen$^{7,8}$,
\newauthor
K.~Ol\'ah$^{9}$,
T.L.~Jacobs$^{10}$,
E.~Newton$^{11,12}$,
M.R.~Omohundro$^{13}$,
D.~LaCourse$^{14}$,
\newauthor
H.M. Schwengeler$^{13}$,
I.A.~Terentev$^{13}$,
D.W.~Latham$^2$,
A.~Bieryla$^2$,
M.~Soares-Furtado$^{15}$,
\newauthor
L.G.~Bouma$^{15}$,
M.J.~Ireland$^{16}$,
J.~Irwin$^{2}$
\\ \\
$^{1}$ Department of Physics, and Kavli Institute for Astrophysics and Space Research, M.I.T., Cambridge, MA 02139, USA; sar@mit.edu \\
$^{2}$ Harvard-Smithsonian Center for Astrophysics, 60 Garden Street, Cambridge, MA 02138 USA \\ 
$^{3}$ Hubble Fellow \\
$^{4}$ Department of Astronomy, The University of Texas at Austin, 2515 Speedway, Stop C1400, Austin, TX 78712 \\
$^{5}$ NASA Sagan Fellow \\
$^{6}$ Department of Physics and Astronomy, University of North Carolina at Chapel Hill, Chapel Hill, NC 27599-3255, USA \\
$^{7}$ DTU Space, National Space Institute, Technical University of Denmark, Elektrovej 327, DK-2800 Lyngby, Denmark \\
$^{8}$ Brorfelde Observatory, Observator Gyldenkernes Vej 7, DK-4340 T\o ll\o se, Denmark \\
$^{9}$ Konkoly Observatory, Research Centre for Astronomy and Earth Sciences, HAS, H-1121 Budapest, Konkoly Thege M. u. 15-17, Hungary\\
$^{10}$ Amateur Astronomer, 12812 SE 69th Place Bellevue, WA 98006 \\
$^{11}$ M.I.T.~Kavli Institute for Astrophysics and Space Research, M.I.T., Cambridge, MA 02139, USA \\
$^{12}$ NSF Astronomy and Astrophysics Postdoctoral Fellow \\
$^{13}$ Citizen Scientist \\ 
$^{14}$ Amateur Astronomer, 7507 52nd Place NE Marysville, WA 98270 \\
$^{15}$ Department of Astrophysical Sciences, Princeton University, NJ 08544, USA \\
$^{16}$ Research School of Astronomy and Astrophysics, Australian National University, Canberra, ACT 2611, Australia \\
}

\date{}

\pubyear{}

\begin{document}
\label{firstpage}
\pagerange{\pageref{firstpage}--\pageref{lastpage}}
\maketitle

\begin{abstract}
We have discovered a young M star of mass $0.16\,M_\odot$ and radius $0.63\,R_\odot$, likely in the Upper Sco Association, that exhibits only a single $80\%$ deep occultation of 1-day duration.  The star has frequent flares and a low-amplitude rotational modulation, but is otherwise quiet over 160 days of cumulative observation during {\em K2} Campaigns C2 and C15.  We discuss how such a deep eclipse is not possible by one star crossing another in any binary or higher-order stellar system in which no mass transfer has occurred.  The two possible explanations we are left with are (1) orbiting dust or small particles (e.g., a disk bound to a smaller orbiting body, or unbound dust that emanates from such a body); or (2) a transient accretion event of dusty material near the corotation radius of the star.  In either case, the time between such occultation events must be longer than $\sim$80 days.  We model a possible orbiting occulter both as a uniform elliptically shaped surface (e.g., an inclined circular disk) and as a `dust sheet' with a gradient of optical depth behind its leading edge.  The required masses in such dust features are then $\gtrsim 3 \times 10^{19}$ g and $\gtrsim 10^{19}$ g, for the two cases, respectively.

\end{abstract}




\section{Introduction}
\label{sec:intro}

In recent years, owing to the advent of precise and continuous photometric monitoring from space, a substantial number of stars exhibiting ``dusty occultations'' have been discovered.  And, it seems, quite a range of physically diverse systems can display such dusty occultations.  The first illustrative class of such objects are the so-called ``dippers'' (see, e.g., \citealt{alencar10}; \citealt{morales11}; \citealt{cody14}; \citealt{ansdell16}; \citealt{hedges18})\footnote{This class of objects was first identified by ground-based observations \citep{bouvier99}.}.  The dips are typically $\sim$10-50\% deep and cannot be attributed to rotational spot modulation or to other intrinsic stellar variability.  These dips can last from hours to days and may be aperiodic, quasi-periodic, or periodic (\citealt{cody14}; \citealt{ansdell16}).  The dips have been associated with occultations by dusty material near the inner edge of circumstellar disks or even corotating with the stellar magnetosphere itself (e.g., \citealt{mcginnis15}). The dippers are mostly young M stars and the ones reported by \citet{ansdell16} show large infrared excess and evidence for debris disks.  Many of the dippers are found in young stellar associations such as Upper Scorpius and the $\rho$ Ophiuchus complexes (\citealt{cody14}; \citealt{ansdell16}; \citealt{hedges18}; \citealt{mcginnis15}).  

A different kind of dust occulter is KIC 8462852 (\citealt{boyajian16}; aka ``Boyajian's star'').  In the discovery paper, there were 10 dips reported over the 4-year interval of the {\em Kepler} observations, ranging in depth between a fraction of a percent and 21\%.  The dips are irregular in shape, can last for days, and are not easily modeled in any simple way. More recent observations firm up the idea that the occultations are due to dusty material \citep{boyajian18}, and suggest that the appearance of the dusty material may be associated with a periodicity of $\sim$1500 days (\citealt{bourne18}; \citealt{boyajian18}).  A key difference between KIC 8462852 and the `dippers' is that the host star in KIC 8462852 is a main-sequence field F star rather than an M star. 

More recently, \citet{rappaport18} reported plausible exocomet transits in two {\em Kepler} stars: KIC 3542116 and KIC 11084727.  The candidate exocomet dips are only $\sim$0.1\% deep, but have a shape that might be expected from dusty trailing comet tails.  

Another category of dusty occultations involves the so-called `disintegrating planets': KIC 1255b, K2-22b, and KOI-2700b (see \citealt{vanlieshout18}, and references therein).  These three objects produce transits that are periodic, but of variable depths, and are thought to be due to dusty material that is being continually emitted by a quite small planet; the hard body transits of the underlying objects themselves have not yet been seen (\citealt{rappaport12};  \citealt{perezbecker13}).

A thus-far unique object exhibiting dusty occultations is WD 1145+017 \citep{vanderburg15}.  This 17th magnitude white dwarf was observed with {\em K2} to exhibit a number of periodicities with periods between 4.5 and 4.8 hours.  The dips had characteristic depths of only $\sim$1\% and duty cycles within these periodicities of $\sim$30\% of the orbit.  The dips are thought to arise from dust clouds anchored or produced by orbiting asteroids (see, e.g., \citealt{vanderburg15}; \citealt{rappaport16}), and numerous ground-based follow-up observations revealed dips of up to 60\% depth (see, e.g., \citealt{rappaport16}; \citealt{gaensicke16}; \citealt{gary17}; \citealt{rappaport18}).

An intriguing object with a deep and long dip is 1SWASP J140747.93-394542.6 (hereafter `J1407'; see \citealt{mamajek12}; \citealt{kenworthy15a}; \citealt{kenworthy15b}).  This star is a member of the Upper Centaurus-Lupus subgroup of the Sco-Cen association of stars (age of 16 Myr).  J1407 undergoes a dip that reaches a depth of $\sim$ 3 magnitudes and lasts for $\sim$45 days.  \citet{kenworthy15b} proposed that the overall long dip seen in the sparsely sampled superWASP photometry was due to a large (0.6 AU) 37-ring system transiting across the host K star.

Another class of dippers are young T-Tauri stars that exhibit month to year long dimming events with depths of up to a couple of  magnitudes (e.g., \citealt{rodriguez17a}; \citealt{rodriguez17b}). Some of these events are interpreted as being due to occultations by tidally disrupted disks due to binary interactions (e.g., \citealt{rodriguez18}). 

Among this list of illustrative occultations by `soft bodies' (i.e., dust) is the following interesting object. \citet{osborn17} report two long dips in the star PDS 110 that are 30\% deep, are roughly `V' shaped, last for $\sim$25 days, and are separated by an 808-day interval.  The observations were made with KELT and WASP photometry.  The 808-d interval between the dips was interpreted as being due to transits by an ``unseen low-mass planet or brown dwarf with a large circum-secondary disk of diameter 0.3 AU'' orbiting at $\sim$2 AU (\citealt{osborn17}; however their originally claimed periodicity has been withdrawn: \citealt{osborn18}). 

Finally, we can add to this list the possible detection of a ringed planet around the M star KIC 10403228 \citep{aizawa17}. These authors studied the transit shapes of 89 {\em Kepler} objects with either a single transit or just a few transits, indicating very long-period planets.  They searched for small asymmetries in the transit profiles and found one suggestive case where a ringed planet fits the transit profile better than a naked planet.

In this work, we report the discovery of a one-day-long, 80\% deep, occultation of a young star in the Upper Scorpius stellar association: EPIC 204376071. Three things that make this very deep and long dip somewhat unique are: (1) the continuous coverage with 1/2-hour sampling of the flux, (2) the very clearly mapped out asymmetry in the occultation profile, and (3) the lack of pronounced WISE 3 and 4 band emission.  In Sect.~\ref{sec:K2} we report on the {\em K2} observations of this target, leading to its discovery.  The available archival data for this object are presented in Sect.~\ref{sec:archive}. Section \ref{sec:spectra} discusses the two ground-based spectra that we obtained for this star.  The target exhibits a distinct stellar rotational modulation, and we present a detailed spot model in Sect.~\ref{sec:spots}.  In Sect.~\ref{sec:model} the large dip in flux is modeled as an occultation by either an inclined, uniform-density disk or, a dust cloud or dust sheet with a gradient of optical depth passing in front of the host star.  In Sect.~\ref{sec:interpret} we offer several different interpretations for the occultation.  Sections \ref{sec:orbits} and \ref{sec:dust} give some constraints on allowed orbits of an orbiting occulter as well as the required dust mass.  In Sect.~\ref{sec:dipper} we discuss the alternate scenario that this occultation is part of the dipper phenomena.  We summarize our findings in Sect.~\ref{sec:summary}.

\vspace{0.6cm}

\begin{figure}[h!]
\begin{center}
\includegraphics[width=0.99\columnwidth]{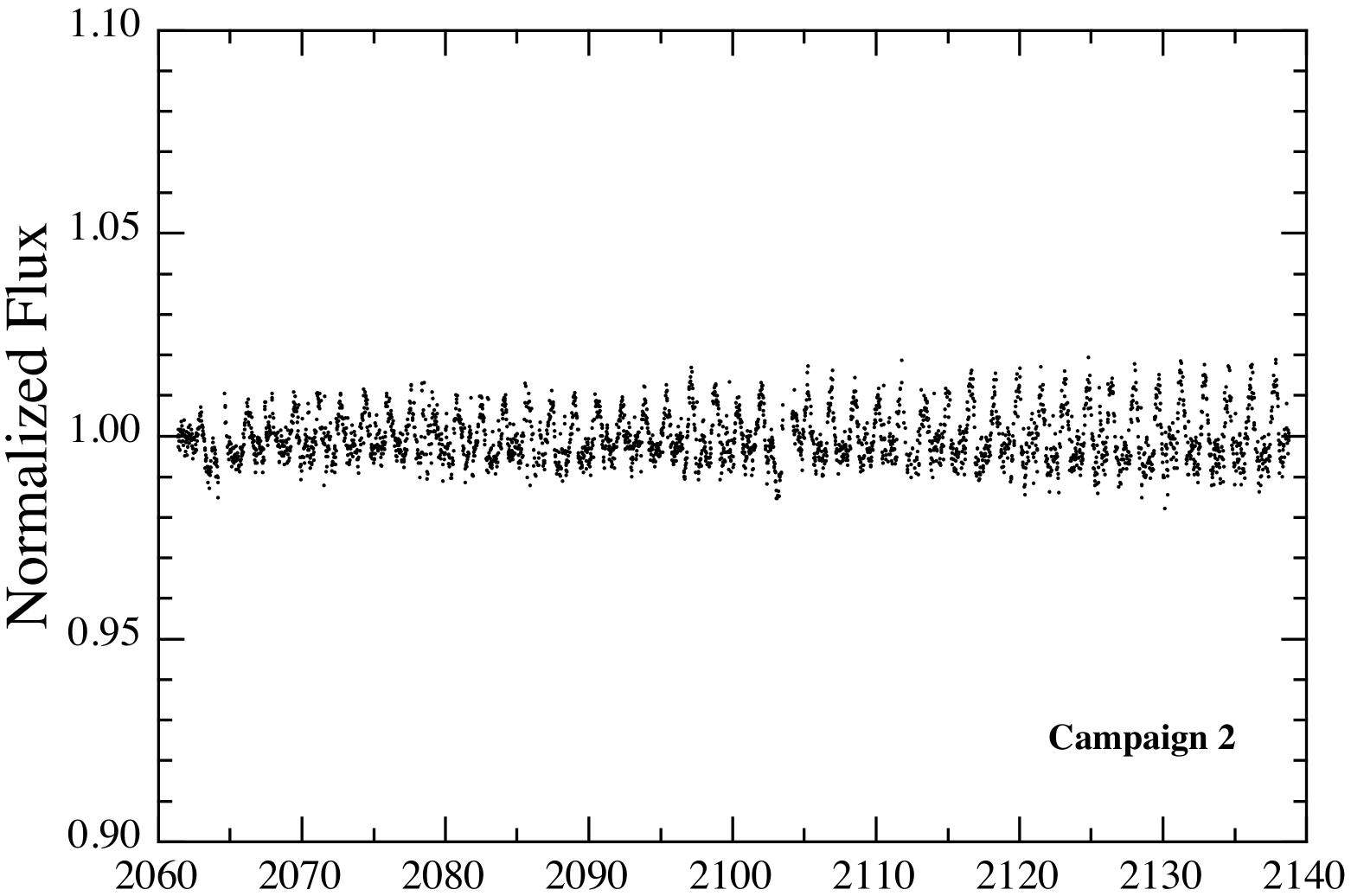} \hglue-0.22cm 
\includegraphics[width=0.97\columnwidth]{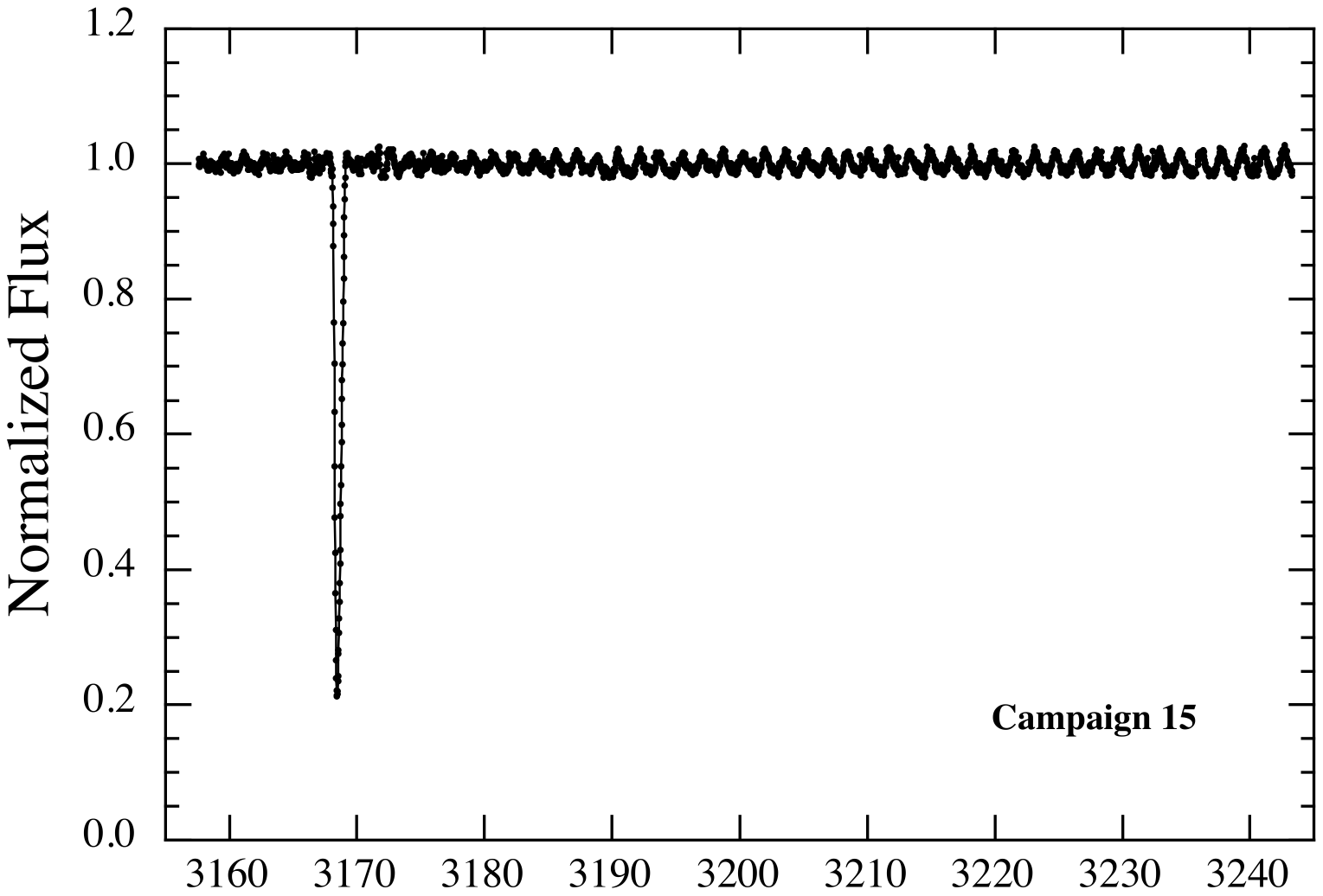} \hglue0.08cm
\includegraphics[width=1.01\columnwidth]{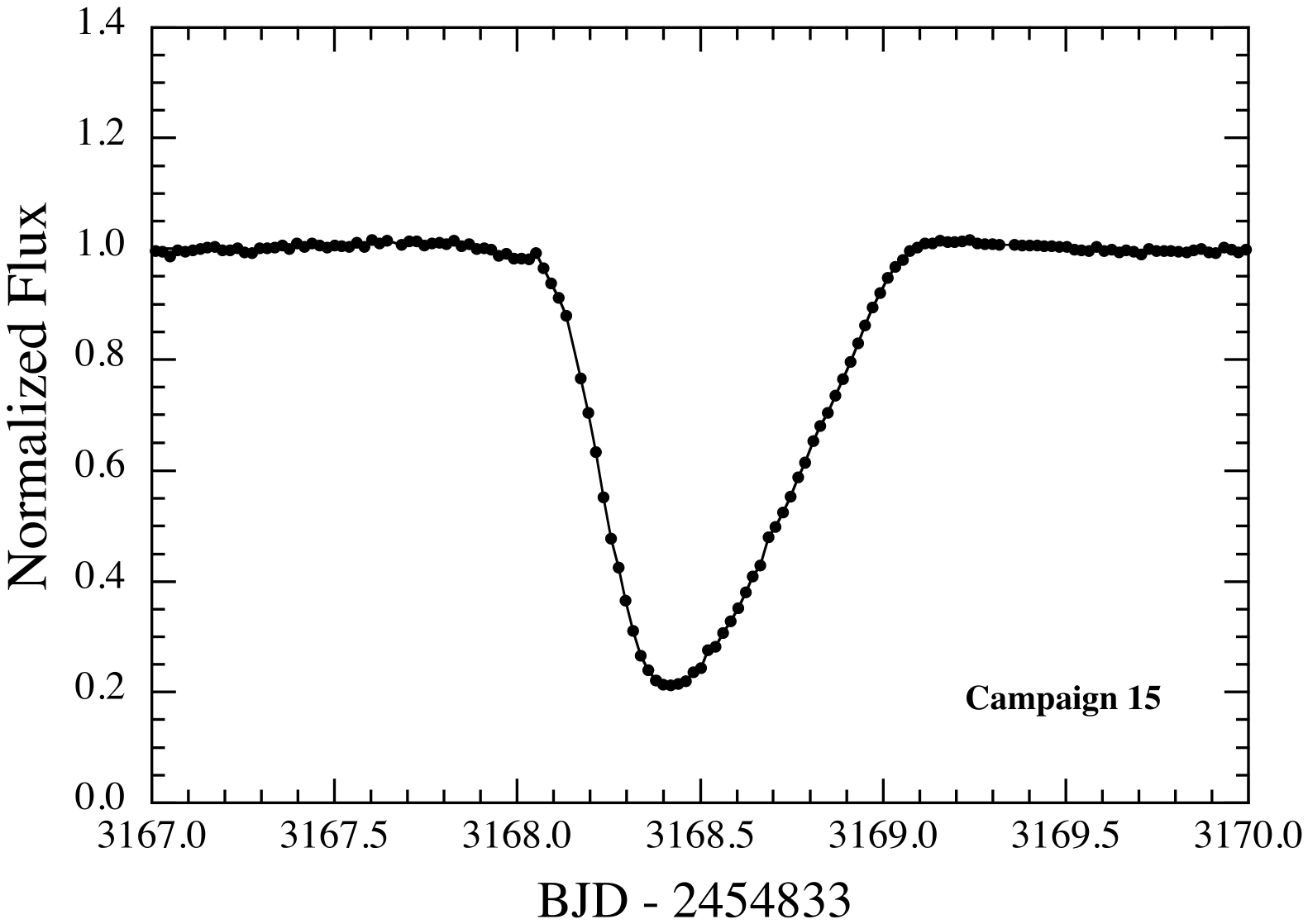}
\caption{K2 lightcurve for EPIC 204376071 during Campaign 15.  Top panel shows the 80-day lightcurve during Campaign 2.  No occultations are seen; the low-amplitude modulation has a period of 1.63 days, which we interpret as a stellar spot rotation period.  Middle panel is the 85-day lightcurve in perspective with a single deep occultation at around KBJD $\equiv$ BJD - 2454833 = 3168. Bottom panel is a zoom-in on the 3 days around the deep occultation, showing that it has a distinctly longer egress than ingress.  Stellar flares have been edited out; see Fig.~\ref{fig:flares}. Note the difference in flux scales between the top panel and the middle and bottom panels.}
\label{fig:rawLC}
\end{center}
\end{figure}  

\section{K2 Observations}
\label{sec:K2}

The {\em Kepler} spacecraft \citep{koch10} observed EPIC 204376071 twice during its $K2$ mission \citep{howell14}, initially in Campaign 2 (C2) (Proposal: GO2063, A.L.~Kraus) and subsequently in Campaign 15 (C15) (Proposals: GO15023, L.A.~Hillenbrand; GO15043, A.C.~Rizzuto). We downloaded the corresponding data from the Barbara A. Mikulski Archive for Space Telescopes (MAST)\footnote{https://archive.stsci.edu/k2/} and created light curves \citep{AV} which were manually scrutinized with the LcTools examination software \citep{kipping15}.  Stand-alone events from $K2$-photometry (e.g., \citealt{zhou18}; \citealt{borko18}) are rarely recovered by automated searches for periodic signals such as the BLS \citep{kovacs02}, unless being coincidentally revealed by additional periodicities from the target star (e.g., EPIC 212813907.01; \citealt{crossfield18}).  Akin to other discoveries of non-repeating events, we detect a single occultation-like event in the C15 light curve of EPIC 204376071 located at 3168.5 BKJD (defined as BJD - 2454833), and blocking up to $\sim$80\% of the light for an entire day. 

After identifying the deep eclipse, we produced a custom light curve for EPIC 204376071 with a handful of modifications from the standard procedures described by \citet{AV}.  First, instead of extracting light curves from stationary apertures, we used a circular moving aperture to reduce the amplitude of systematics due to aperture losses for this particularly faint target.  We allowed our procedure to model higher-frequency stellar variability than we typically do, using a spline with break points spaced every 0.4 days instead of our typically used 1.5-day spacing.  We also increased the length of the ``divisions'' of the light curve on which we perform our one-dimensional systematics correction, which yielded a better correction near the long, deep occultation.  The resulting light curves are shown in Fig.~\ref{fig:rawLC} and used throughout the rest of this paper in our analysis.

The full C2 lightcurve is shown in the top panel of Fig.~\ref{fig:rawLC}.  Aside from an unremarkable 1.6-day periodicity (see Sect.~\ref{sec:spots}; \citealt{rebull18}) with a $\sim$1\% amplitude the flux is rather constant and without any occultations.  The full C15 lightcurve is shown in the middle panel of Fig.~\ref{fig:rawLC} where there is an obvious single large drop in flux about 10 days into the campaign.  The bottom panel of  Fig.~\ref{fig:rawLC} shows a zoom-in on 3 days of the lightcurve centered on the occultation.  Here one can see that the depressed flux lasts for about a day, is extremely deep, and is quite noticeably asymmetric with an egress that is about twice as long as the ingress\footnote{Here we use the terms `ingress' and `egress' even though we do not know for sure if this is an orbiting occulting object (solid or otherwise) in the conventional sense.}.  

The lack of any occultations in the C2 light curve indicates a lower limit to any period for the occultations of $P \gtrsim$ 79 days. Solely using the two K2 observations, separated by almost exactly 3 years (C2 ranging from 23 Aug 2014 - 10 Nov 2014 and C15 ranging from 23 Aug 2017 - 19 Nov 2017), we can further constrain any allowed periods. We simply note here the first five allowed windows in period for a potentially orbiting body: $P \simeq 79$ d, 85 d, 93 d,  101-103 d, and 111-114 d.  

The astrophysical nature of the occultation event is supported by a target pixel file animation of the event, using custom software similar to {\tt K2Flix}: a {\em Kepler} pixel data visualization tool\footnote{\url{http://barentsen.github.io/k2flix/}}, in which EPIC 204376071 clearly dims dramatically\footnote{We have included the video in MP4 format in the on-line-only supplementary material for this paper; the file is named `occultation.ep204376071.mp4'.  An expression for converting time to frame number, $f$, is $f \simeq 585 + 48.94 \times ({\rm KBJD} - 3168.4)$, where frame 585 is near the bottom of the occultation.} and thereby indicates that a data anomaly is not the explanation for the dip.  To further ensure the astrophysical origin of this occultation event, we checked for systematics and found a significant centroid shift of $\simeq 0.3$ pixels during the occultation in the direction transverse to $K2$'s roll. We believe this is an artefact of the backgrounds; when the flux of the star decreases by a factor of 5, if there is any uneven background, that could introduce a shift this large. Finally, we noticed a photometric peak near the occultation event at BKJD = 3171.7, with a 40\% {\em rise} in flux, but this is clearly identified as an inner main belt contaminant `Hawke (3452)' using {\tt SkyBot} \citep{berthier16}, and this peak was removed.

\begin{table}
\centering
\caption{Properties of the EPIC 204376071 System}
\begin{tabular}{lc}
\hline
\hline
RA (J2000) & 16:04:10.122   \\  
Dec (J2000) &  $-$22:34:45.36  \\  
$K_p$ & 14.91  \\
$G$$^a$ & $16.055 \pm 0.001$ \\
$G_{\rm rp}$$^a$ & $14.678 \pm 0.002$ \\
$G_{\rm bp}$$^a$ & $18.132 \pm 0.012$ \\
J$^b$ & $12.47 \pm 0.02$  \\
H$^b$ & $11.84 \pm 0.02$ \\
K$^b$ & $11.55 \pm 0.03$  \\
W1$^c$ & $11.34 \pm 0.02$ \\
W2$^c$ & $11.08 \pm 0.02$ \\
W3$^c$ & $10.70 \pm 0.11$  \\
W4$^c$ & $\gtrsim 8.6$  \\
$T_{\rm eff}$ (K)$^a$ & $3861 \pm 450$\\
Distance (pc)$^a$ & $135.3 \pm 3.6$ \\   
$\mu_\alpha$ (mas ~${\rm yr}^{-1}$)$^a$ & $-11.54 \pm 0.43$  \\ 
$\mu_\delta$ (mas ~${\rm yr}^{-1}$)$^a$ &  $-24.89 \pm 0.18$  \\ 
\hline
\label{tbl:mags}
\end{tabular}

{\bf Notes.} (a) Gaia DR2 \citep{lindegren18}.  (b) 2MASS catalog \citep{Skrutskie}.  (c) WISE point source catalog \citep{cutri13}.  
\end{table}

\vspace{0.5cm}

\section{Archival Data}
\label{sec:archive}

We have collected the available photometric data on the target-star in Table \ref{tbl:mags}.  

Gaia DR2 places the object at a distance of $135 \pm 4$ pc \citep{lindegren18}.  They also list a much fainter nearby star that is $2''$ to the south of the target star and has $\Delta G = 4.26$.  This faint companion is likely physically unrelated to the target star because the proper motions of the two stars in RA and Dec differ  by $11.7 \pm 3.5$ and $20.6 \pm 1.5$ mas/year, respectively.  

We were able to use the PanSTARRS $g$ and $r$ band images \citep{chambers16} to show that the faint companion star has 8\% and 5\% of the fluxes of the target star, respectively. By contrast, the faint companion has only 2\% of the Gaia broad band flux.  Thus, we take the occultation depth of the target star in the {\em Kepler} band to be increased from $\simeq78\%$ to $\simeq 80\%$ after taking into account the small dilution effect of the faint neighbor star. 

The distance, proper motion, and radial velocity suggest that EPIC 204376071 is a member of the Upper Scorpius association of stars,  which is a subgroup of the Scorpius-Centaurus OB association. The age of Sco-Cen has historically been assumed to be approximately 5 Myr \citep{degeus89}, but has recently been suggested to be $11 \pm 3$ Myr \citep{pecaut16}; we adopt the latter age in this work.  Some of the properties of the Upper Sco group, in particular the location on the sky, the distance, and the common proper motion are summarized in Table \ref{tbl:UpSco}, along with the same attributes for the target star.  As is readily apparent, the target star sits well within the Upper Sco association in position and velocity space.  We have also utilized the online tool {\tt BANYAN $\Sigma$}\footnote{\url{http://www.exoplanetes.umontreal.ca/banyan/banyansigma.php}}, the multivariate Bayesian algorithm that identifies membership in young stellar associations out to 150 pc \citep{gagne18}.  The probability that EPIC 204376071 is a member of the Upper Scorpius association assigned by {\tt BANYAN} $\Sigma$ is 99.9\%, which is a strong indication of membership.  Therefore, from hereon out we will assume that EPIC 204376071 is indeed a member of Upper Sco and therefore has an age of $\sim$10 Myr (\citealt{pecaut12}; \citealt{feiden16}).

This adopted age seems consistent with the inferred rotation period of 1.63 days \citep{irwin11,mcquillan13,reinhold13,rebull18}. While rapid rotation can be caused by tidal synchronization with a close binary companion \citep{simonian18}, we currently have no photometric or spectroscopic indications that EPIC 204376071 is a binary.

\begin{figure}
\begin{center}
\includegraphics[width=1.00 \columnwidth]{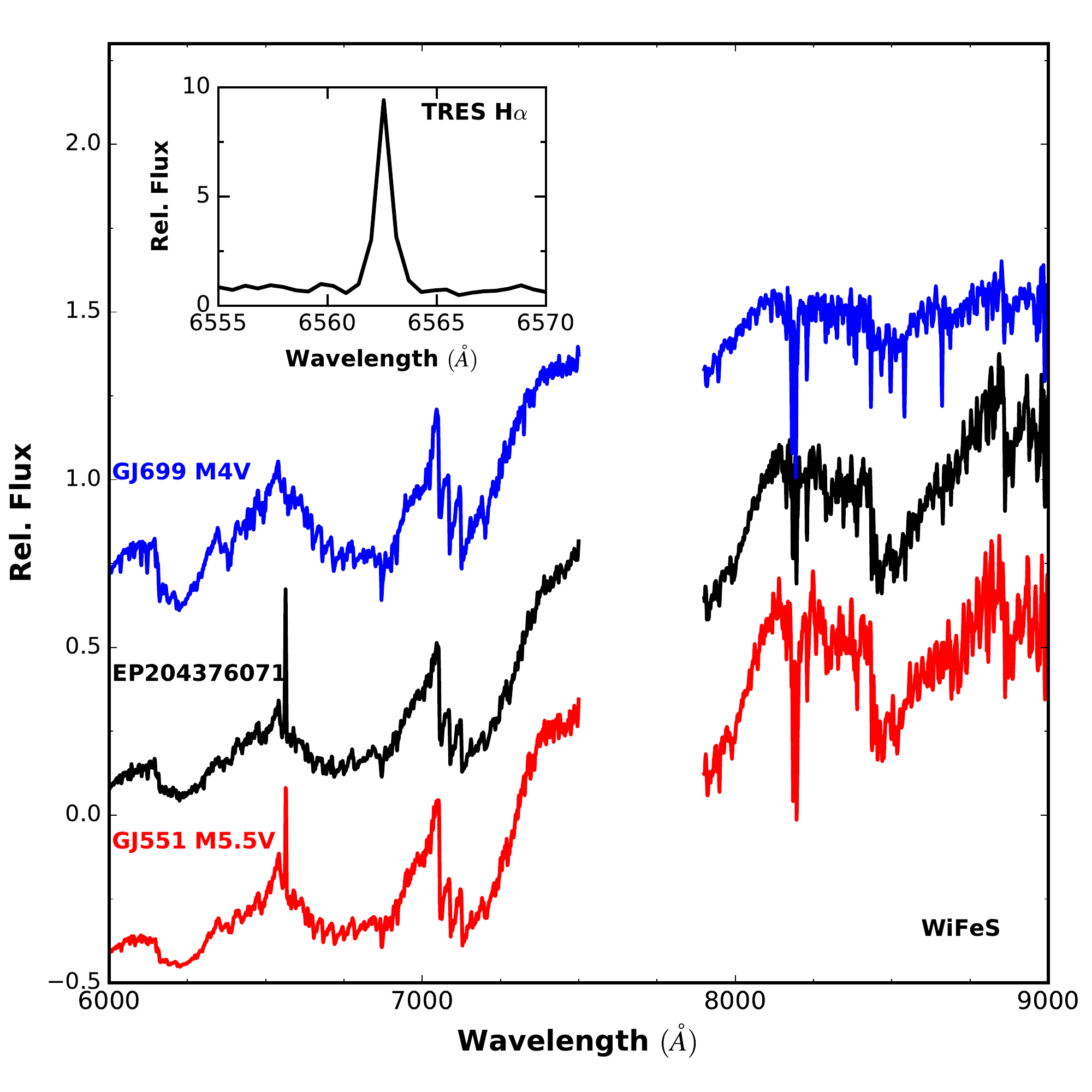}
\caption{Low resolution spectrum of EPIC 204376071 from the ANU WiFeS spectrograph (black; see Sec.\ref{sec:spectra}), compared against the spectra of Barnard's star (M4, blue) and Proxima Centauri (M5.5, red). {\em Inset:} The TRES spectrum over the H$\alpha$ region showing the strong single-cored emission feature.}  
\label{fig:spectra}
\end{center}
\end{figure}  

We have checked various other archival data bases (e.g., DASCH, \citealt{grindlay09}; HATSouth, \citealt{bakos13}; SWASP, \citealt{pollacco06}; ASAS, \citealt{pojmanski97}; ASAS-SN, \citealt{shappee14}) searching for additional occultations.  However, we found that the source was either not observed or is sufficiently faint that only unusable upper limits were obtained.

\section{Spectra}
\label{sec:spectra}

\begin{table}
\centering
\caption{Location of EPIC 204376071 within Upper Scorpius}
\begin{tabular}{lccc}
\hline
  & 204376071 & Upper Sco$^a$ & Range$^a$ \\
       &   & (mean) & (half width)   \\
\hline
RA (J2000) [deg] & 241.04  & 241.0 & 4 \\
Dec (J2000) [deg]& $-$22.58 & $-$21.8 & 4 \\
d (pc) & $135 \pm 4$$^b$ & 132 & 17 \\
RV [km s$^{-1}$] & $-8.7^c$ & $-5$ & 4 \\ 
$\mu_\alpha$ (mas ~${\rm yr}^{-1}$) & $-$11.5$^b$  & $-11$ & 7 \\
$\mu_\delta$ (mas ~${\rm yr}^{-1}$) & $-24.9^b$ & $-23$ &  7 \\
age (Myr) & ... & $11 \pm 3$$^d$ &  ... \\
\hline
\label{tbl:UpSco}
\end{tabular}

{\bf Notes.} (a) \citet{gagne18}. (b) From the Gaia DR2 results \citep{lindegren18}.  (c) From the TRES spectrum (see Sect.~\ref{sec:spectra}). (d) \citet{pecaut12}. 
\end{table}

We obtained low- and high-resolution spectra of EPIC 204376071 to estimate its stellar properties and activity level. We made use of the Wide Field Spectrograph \citep[WiFeS;][]{dopita07} on the ANU 2.3\,m telescope at Siding Spring Observatory in Australia. The WiFeS spectra were obtained at a spectral resolution of $\lambda / \Delta \lambda \equiv R = 3000$, over the wavelength range of 6000 - 9000 \AA, and the spectral extraction and flux calibration follow the procedures laid out in \citet{bayliss13}. The WiFeS spectrum of EPIC 204376071 is consistent with that of an M5 star, and is compared against that of Barnard's star (GJ 699) and Proxima Centauri (GJ 551) in Figure~\ref{fig:spectra}. 

A spectrum of EPIC 204376071 was also obtained with the Tillinghast Reflector Echelle Spectrograph (TRES; \citealt{tres07}; \citealt{furesz08}) on the 1.5\,m telescope at Fred Lawrence Whipple Observatory in Arizona, USA. TRES is a fibre fed spectrograph that yields a spectral resolution of $R=44000$. Due to the faintness of the target star, our TRES spectrum (with 3600 s exposure) is of low signal-to-noise over most of the spectral range. However, we were able to identify strong single-lined emissions over the Balmer series, with the H$\alpha$ emission having an equivalent width of 14 \AA, and derive a barycentric velocity for the star of $-8.7$ km s$^{-1}$ using the TiO lines at $\sim$7100  \AA. An examination of the same spectral region yields an estimated stellar rotational broadening velocity of $v \sin i \simeq 17 \, \mathrm{km\,s}^{-1}$, though the uncertainties in the measurement are difficult to derive due to the low SNR nature of the spectrum. The H$\alpha$ emission as seen from TRES is shown in the inset panel in Figure~\ref{fig:spectra}.

\vspace{0.5cm}

\begin{figure}
\begin{center}
\includegraphics[width=0.99 \columnwidth]{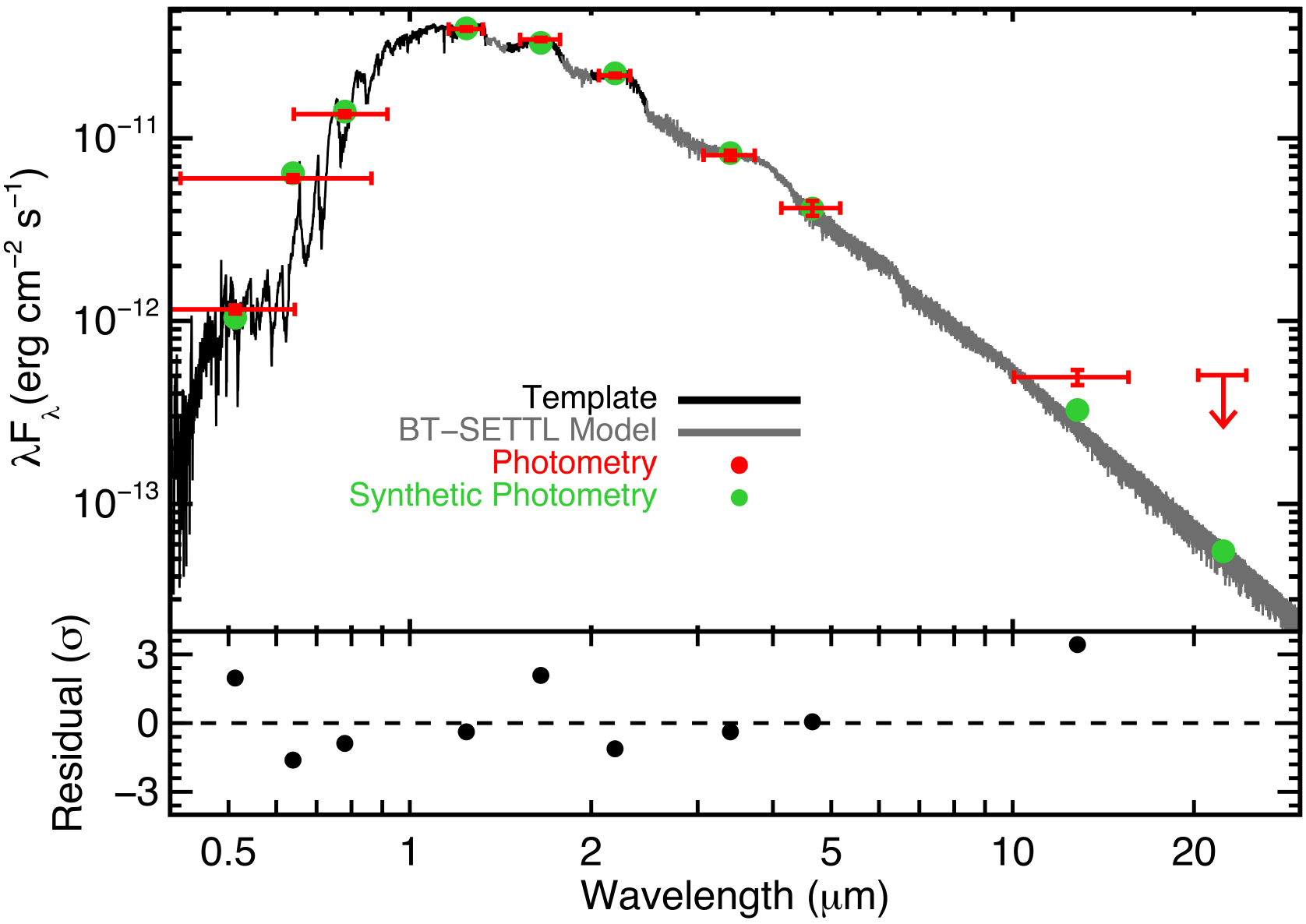}
\caption{SED of EPIC 204376071, with the optical defined by the Gaia $G_\mathrm{BP}$ and $G_\mathrm{RP}$, and the infrared by the 2MASS and WISE magnitudes. Observed magnitudes are shown in red, with vertical errors corresponding to the observed magnitude uncertainties (including stellar variability) and the horizontal errors corresponding to the width of the filter. Synthetic magnitudes from the best-fit template spectrum (black) and BT-SETTL model (grey) are shown as green circles.}
\label{fig:sed}
\end{center}
\end{figure}

To determine the physical parameters of EPIC 204376071 (luminosity, radius, and $T_\mathrm{eff}$) we used the observed spectral energy distribution (SED). We made use of photometric magnitudes from Gaia $G_\mathrm{BP}$ and $G_\mathrm{RP}$ \citep{gaia18}, 2MASS $J$, $H$, $Ks$ bands \citep{cutri03}, and WISE $W1$, $W2$, $W3$ \citep{cutri13} to describe the SED of EPIC 204376071 (Figure~\ref{fig:sed} and Table~\ref{tbl:mags}). 

\begin{table}
\centering
\caption{Properties of EPIC 204376071$^a$}
\begin{tabular}{lc}
\hline
$E(B-V)$ & $0.150 \pm 0.031$ \\
$L$ [L$_\odot$] & $0.0273 \pm 0.0020$ \\
$R$ [R$_\odot$]  & $0.631 \pm 0.042$ \\
$M$ [M$_\odot$] & $0.161 \pm 0.028$ \\
$T_{\rm eff}$ [K] & $2960 \pm 75$ \\
\hline
\label{tbl:parms}
\end{tabular}

{\bf Notes.} (a) Deduced from the analysis of the spectrum and the SED (see Sect.~\ref{sec:spectra}).
\end{table}

We then compared this SED to a set of un-reddened optical and near-infrared spectra from the TW Hydra or $\beta$ Pic moving groups \citep[10-25\,Myr,][]{bell15}, following the method from \citep{mann16}, which we briefly summarize here. We restricted our comparison to templates with spectral types consistent with EPIC 204376071, which we determined to be M5.3$\pm$0.7 by comparing TiO and CaH indices from \citet{lepine13} in the observed WiFeS spectrum to those measured from the young spectra in \citet{herczeg14}. Spectral types for the SED-fitting templates were determined in an identical manner, and hence were on the same overall scale, and depend only weakly on the overall reddening. Gaps in the coverage of each template were filled with PHOENIX BT-SETTL models \citep{allard12}. We inflated the uncertainties on Gaia magnitudes by 2-3\% to reflect stellar variability observed in the \textit{Kepler} bandpass. No template could reproduce the observed $W3$ magnitude and Gaia optical photometry simultaneously, possibly due to the presence of dust/debris past a few AU\footnote{It is difficult to say anything meaningful about a possible large-scale dust/debris structure with a single data point showing a small excess. There are multiple free parameters (dust temperature, mass, opacity), and insufficient information to constrain them. Moreover, the W3 excess is only marginally significant, and thus we prefer to avoid over-interpreting this single data point.}, so we excluded it from the fit. 

For each template, we found the best-fit reddening value ($R_V$ assumed to be 3.1) to reproduce the observed photometry, and integrated over the un-reddened template to determine the total bolometric flux. We calculated the luminosity ($L_*$) using the bolometric flux and the Gaia-measured parallax. Each template had a $T_\mathrm{eff}$ assigned based on the fitting method described in \citep{mann13}, which reproduces empirical $T_\mathrm{eff}$ estimates from long-baseline optical interferometry \citep{boyajian12, mann15} and young eclipsing binaries \citep{Kraus2015}. We computed the corresponding radius ($R_*$) for each template using our assigned $L_*$ Stefan-Boltzmann relation ($L_*=4\pi\sigma R_*^2T_\mathrm{eff}^4$). Lastly, we determined a stellar mass ($M_*$) by interpolating the estimated $L_*$ and $T_\mathrm{eff}$ onto a grid of magnetic models optimized for the age of Upper Scorpius \citep{feiden16}. 

Final uncertainties in each parameter account for the range of possible template fits (each of which yields a slightly different $T_\mathrm{eff}$, E(B-V), luminosity, and radius), as well as measurement uncertainties in the templates and SED. We also took into account errors introduced from filling gaps with models, calibration errors in the $T_\mathrm{eff}$ scale, and uncertainties in the photometric zero-points. The $M_*$ uncertainties also accounted for a range of possible ages in Upper Scorpius due to age spreads ($\pm$5\,Myr), although in practice, ages more $\simeq$3\,Myr off were inconsistent with either the assigned $L_*$, $T_\mathrm{eff}$, or un-reddened photometry, and hence were down-weighted in the analysis. 

These properties of EPIC 204376071 are summarized in Table \ref{tbl:parms}.

The historical definition for the boundary between weak-lined and classical T Tauri stars of a 10-\AA~equivalent width has been updated numerous times to apply for later spectral types. For example, \citep{white03} define an H-$\alpha$ equivalent width boundary of 20 \AA~for M3 to M5.5 spectral types, which places EPIC 204376071 as a weak lined T~Tauri star consistent with no accretion. This is also consistent with the lack of an inner disk excess in the W1 and W2 bands. The WISE W3 excess of $0.47 \pm 0.15$ magnitudes appears to be genuine, as there is no nearby background source in the WISE images that could cause the excess. EPIC 204376071 then arguably has a debris disk, with the W4 upper limit putting it outside the typical excess range of late-type transitional disks \citep{cieza12}.

\section{Flares and Spots}
\label{sec:spots}

The 1.6-day periodic flux modulation of EPIC~204376071 (see also \citealt{rebull18}) is very stable, indicating long-lived active regions on its surface. Its mass, about 0.16\,M$_\odot$, is well below the fully-convective limit of 0.30-0.35\,M$_\odot$ \citep{chabrier97}. Stable spot configurations are typical of low mass, fully convective stars. A good example is the M4 dwarf V374~Peg which has a mass of 0.30\,M$_\odot$, just below the fully-convective limit, has a rotation period that is close to 0.5 days, and has a stable light curve for 16 years with only minor changes \citep{vida16}.  

Rapidly rotating, fully convective M-dwarfs can show spots at all latitudes, as were found on two stars via Doppler imaging \citep{barnes15}. One of the two stars, the primary of a wide binary, the M4.5-dwarf GJ791.2A, rotates faster (0.31 days) and is a bit more massive (0.29\,M$_\odot$) than EPIC~204376071. It has a photospheric temperature of 3000~K, the same (within the errors) as our star, and from the Doppler maps \citet{barnes15} found the temperature difference between the photosphere and spots to be $\Delta T=300$~K. 

\begin{figure}
\begin{center}
\includegraphics[width=1.00 \columnwidth]{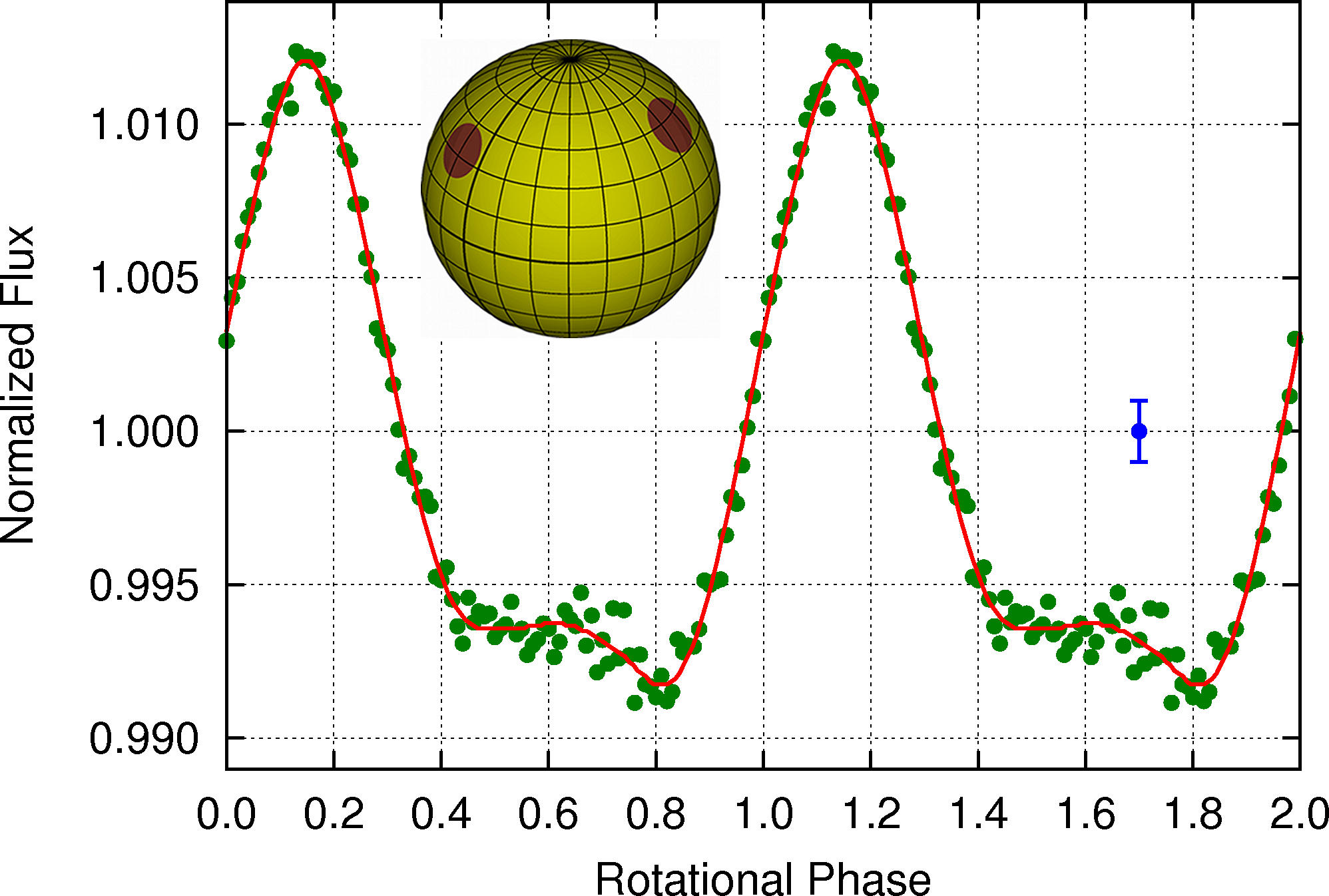}  
\caption{The K2 lightcurve of the low-amplitude modulations in EPIC 204376071 during C2 and C15 folded about a period of 1.629 days. Green points are the folded data; smooth red curve is the fit to a two-spot model (see text for details).  The inset is a visual representation of the spot model. The rms scatter per bin in the fold is close to 0.005 while the formal standard error per bin is 0.0006; the blue error bar (0.001) reflects the empirical rms scatter about the model fit.}
\label{fig:rotation}
\end{center}
\end{figure}  

The ``spot temperature'' signifies a temperature mixture of cool (spots) and hot (faculae) regions, which cannot be resolved. Therefore the area on the stellar surface covered by spots in the models shows up with the average temperature of the hot and cool regions. The very active M dwarfs have numerous hot regions in addition to the cool spots. In the case of V374~Peg, the spot properties result from 4-color photometry, and the spot temperature value is just 150~K below the photosphere \citep{vida16}.

\begin{figure*}
\begin{center}
\includegraphics[width=0.80\textwidth]{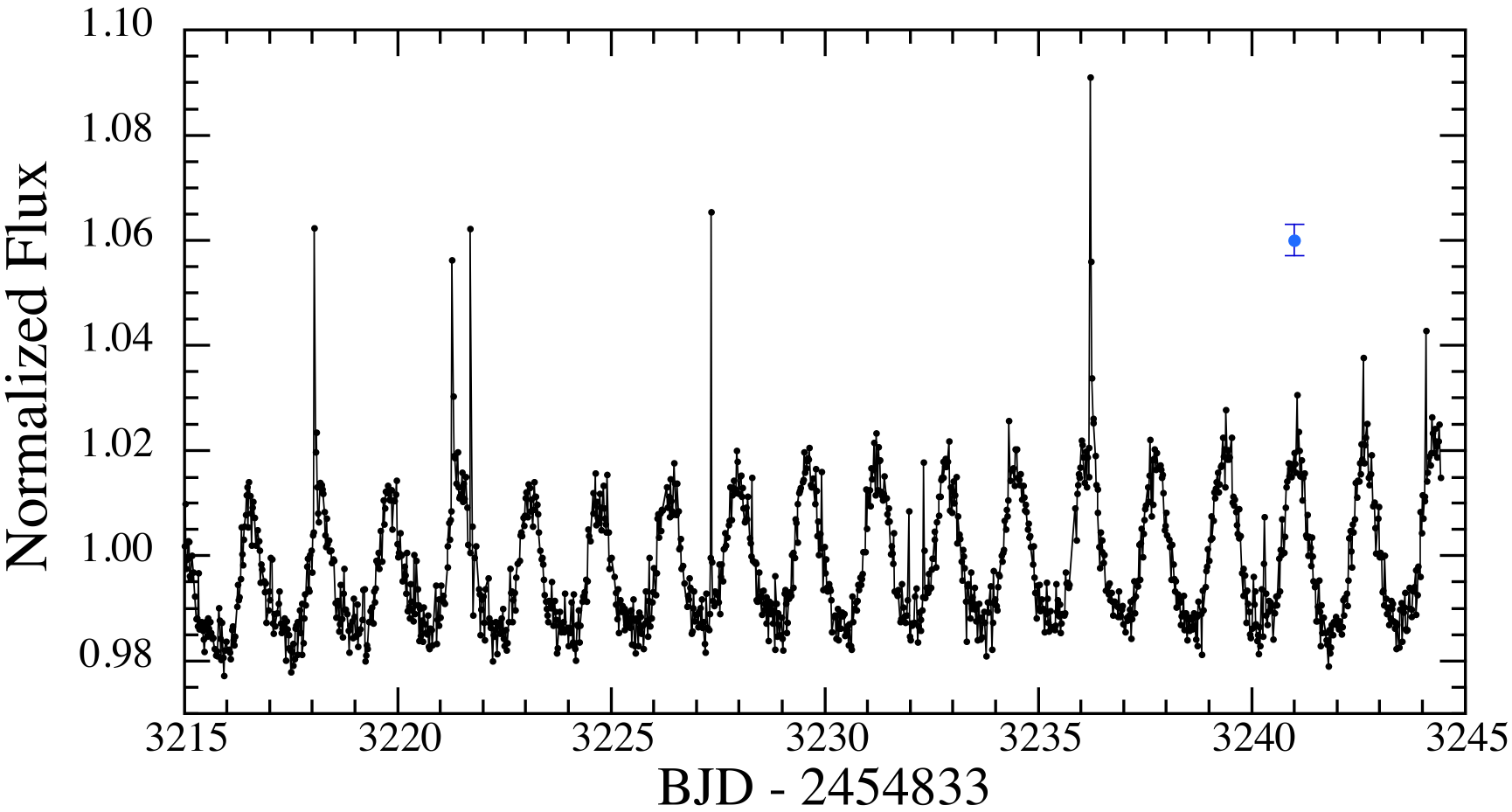}
\caption{A zoom-in on 30 days of the C15 data showing a dozen significant flares, in addition to the 1.63-d rotational modulation.  The same rate of flaring persists throughout C2 and C15. The statistical uncertainties per data point are indicated by the blue point with error bar.}
\label{fig:flares}
\end{center}
\end{figure*}  

We produced an average rotational light curve from all the $K2$ observations of EPIC~204376071 (Fig.~\ref{fig:rotation}) and fitted it \citep{ribarik03} with a simple two-spot model consisting of two circular spots. On the basis of the known spot temperatures of other, fully convective stars \citep{vida16, barnes15} we adopted a spot temperature that is cooler than the photospheric temperature of $\approx 3000$~K by $\sim$200~K. The best fit was found for an inclination angle of $60^\circ$. The resulting fit yields spots that are of similar size at medium latitudes ($\approx 30^\circ$-$40^\circ$) on the star and separated by $\sim$130$^\circ$ in longitude. The spot model configuration for EPIC 204376071 is shown as an inset to Fig.~\ref{fig:rotation}. We note, that the model accounts for only those spots which cause the rotational modulation itself. There may well be more spots on the stellar surface, at the poles and/or evenly distributed, and in this way not causing any additional rotational modulations in flux. Since the unspotted brightness of the star is not known, the total spotted area on the star cannot be estimated.

Acceptable fits to our average light curve can be found from high to medium inclinations and with spot temperature differences with respect to the photosphere of 100~K up to 1000~K, due to the lack of color-dependent rotational lightcurves which can constrain the spot temperature. The uncertainties in the rotational inclination angle and the spot temperature directly affect the latitudes and spot sizes as well. In summary, due to the uncertain inclination angle and spot temperature, and the unknown unspotted brightness, the resulting spot configuration (except the longitudes which follow directly from the observations) is not well constrained, but still shows that with cool spots the rotational light variation can be well modeled.

Besides spots, flares are another typical feature of active M dwarf stars, and these are also found in the lightcurve of EPIC 204376071, as for all of the stars studied by \citet{morin08}, for example.  Those objects were well-known flare stars well before magnetic modeling of their surface came into prominence. An illustrative set of flares from EPIC 204376071 is shown in Fig.~\ref{fig:flares}, with flare amplitudes up to $\approx 6\%$. 

The mass of EPIC~204376071 is $\simeq$ 0.16\,M$_\odot$ whereas its radius is 0.63\,R$_\odot$ indicating that the star has not reached the main sequence. Stars with such a low mass remain fully convective during their entire evolution. Standard evolutionary models would yield a younger age for a star with this mass and radius than that of the Upper Sco association as a whole. However, including magnetic fields in the evolutionary models can resolve this discrepancy, allowing the low-mass stars to be coeval with the rest of the stars in Upper Sco (see \citealt{feiden16}; \citealt{macdonald17}).  \citet{macdonald17} have tested this scenario using an eclipsing binary in Upper Sco, with two low-mass components of 0.12\,M$_\odot$ and  0.11\,M$_\odot$ with radii of 0.42\,R$_\odot$ and 0.45\,R$_\odot$, respectively.  The precise stellar parameters of an eclipsing binary make the comparison to the models more reliable.  The results show that radial magnetic fields of a few hundred Gauss would make the age of these stars consistent with that of Upper Sco. Adopting the results of \citet{macdonald17} we can state that EPIC~204376071 is a low mass, magnetically active star in Upper Sco, with an age of about $11 \pm 3$ Myr reported by \citet{pecaut12}.

\section{Model Fits to the Occultation}
\label{sec:model}

\subsection{General Considerations}
\label{sec:general}

We believe that the deep asymmetric occultation of EPIC 204376071 is likely caused by dust obscuration.  However, we first discuss, and then discard, the idea that this may be some kind of hard-body eclipse by one star blocking the light of another.  Consider a binary system with constituent main-sequence (`MS') stars of masses $\lesssim 2 \,M_\odot$.  The largest fraction of blocked system light occurs during the eclipse of the more massive star, assuming an equatorial eclipse.  When the two MS stars are of comparable mass, that maximum fraction is $\sim$50\%.  If one of the stars is somewhat evolved, then an even smaller fraction of the system light can be eclipsed. If we add more complicated structure to the system by considering triple and quadruple star systems\footnote{In this logical progression, we have skipped over the fact that, of course, two stars in a binary cannot produce such an asymmetric occultation as is observed in EPIC 204376071.  However, as soon as three-body systems are invoked, quite exotically shaped 3rd body eclipses are possible.}, this only makes it harder to block more than 50\% of the system light via eclipses due to the dilution effect of the non-eclipsing stars.  Finally, if the stars have undergone mass transfer then the result may be an Algol-like system where the mass-losing star might be both larger and less luminous than the accreting star, thereby allowing for eclipses $>50\%$.  However, the EPIC 204376071 system appears to house a dominant M star of bloated radius (i.e., not having yet settled onto the main sequence), and therefore we are likely safe in assuming that this is the primordial system, unaltered by mass transfer.   

We therefore invoke two models involving dust: an inclined disk producing an elliptical shape in projection, and a dust cloud or sheet (hereafter `dust sheet') whose geometry need not be defined very precisely.  We note here that a small amount of dust can remove a sizable fraction of the system light without requiring a large mass in dust---we quantify this statement below.  

\subsection{Disk Occulter Model}

The first model for the occulter that we consider is an intrinsically circular disk of uniform optical depth.  Presumably such a disk is orbiting another `anchor' body in the system such as a brown dwarf or large planet.  The orientation of the disk can be described for our purposes simply by first specifying the apparent semiminor axis of the projected disk, which is just the semimajor axis multiplied by $\cos\,i$, where $i$ is the usual inclination angle. The second angle which specifies the disk's orientation will be the angle that the semimajor axis makes with the orbital plane, hereafter the `tilt' angle.  

There are eight free parameters to be fit: the semimajor and semiminor axes; the tilt angle;  $v_t$, the transverse speed of the disk across the host star; $b$, the impact parameter; $\tau$, the optical depth of the disk; $t_0$, the time of closest approach between the disk and host star; and DC, the out-of-occultation background level.  We utilize a Markov Chain Monte Carlo (`MCMC') code (\citealt{ford05}; \citealt{madhu09}; \citealt{rappaport17}) to fit this disk model to the deep occultation event in EPIC 204376071. For each choice of parameters, we generate a model lightcurve by integrating the over the stellar disk, including the disk transmission and limb darkening, and then repeating this in increments of 10 minutes as the disk crosses the stellar disk. The model lightcurve is then convolved with the $\sim$30 minute integration time of the {\em Kepler} long-cadence sampling. Each model is then evaluated via the $\chi^2$ value of the fit to the data, and the code then decides, via the Metropolis-Hastings algorithm, whether to accept the new set of parameters or to try again.  Each MCMC chain has $10^5$ links, and we have run a half-dozen chains.  

\begin{figure}
\begin{center}
\includegraphics[width=1.00 \columnwidth]{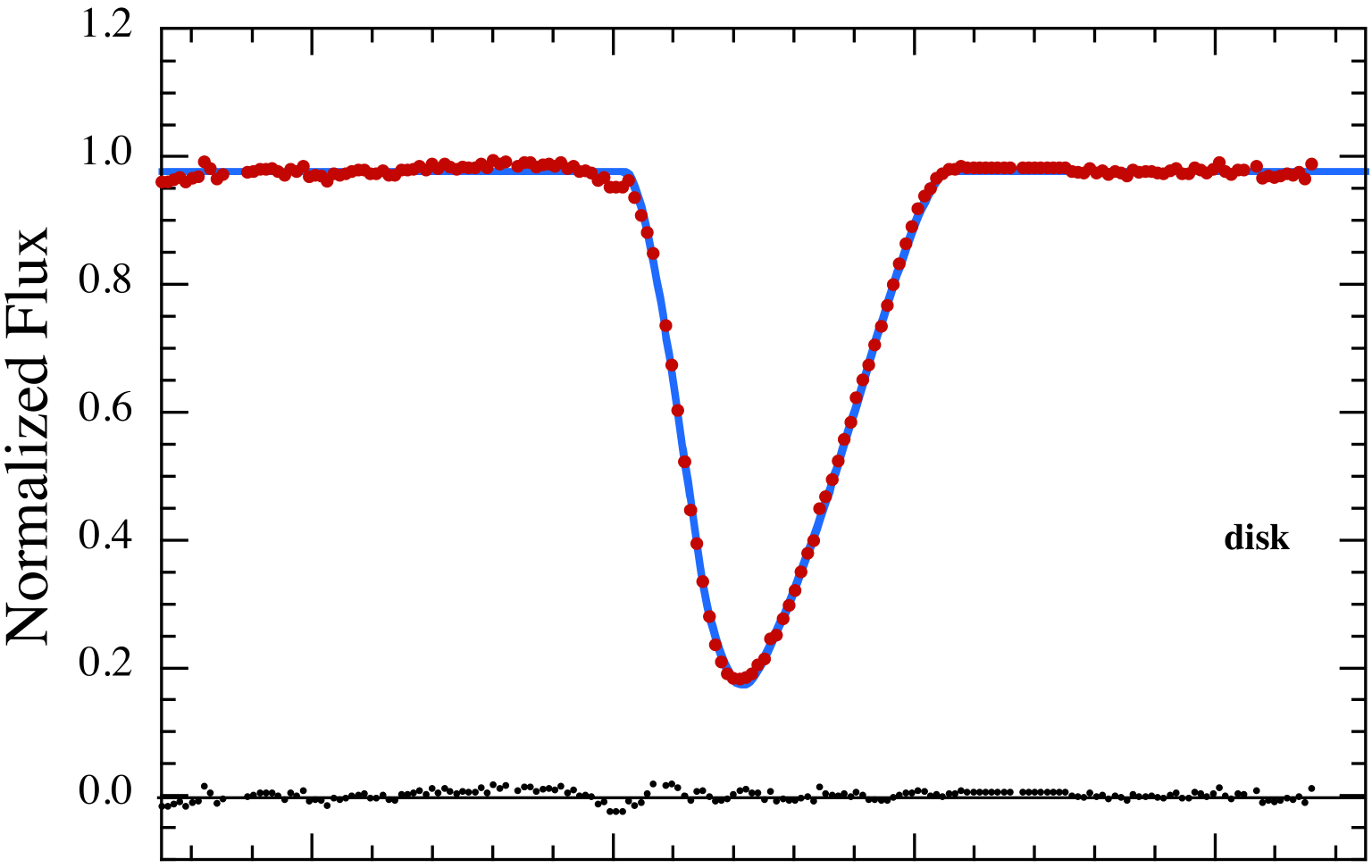}
\includegraphics[width=1.00 \columnwidth]{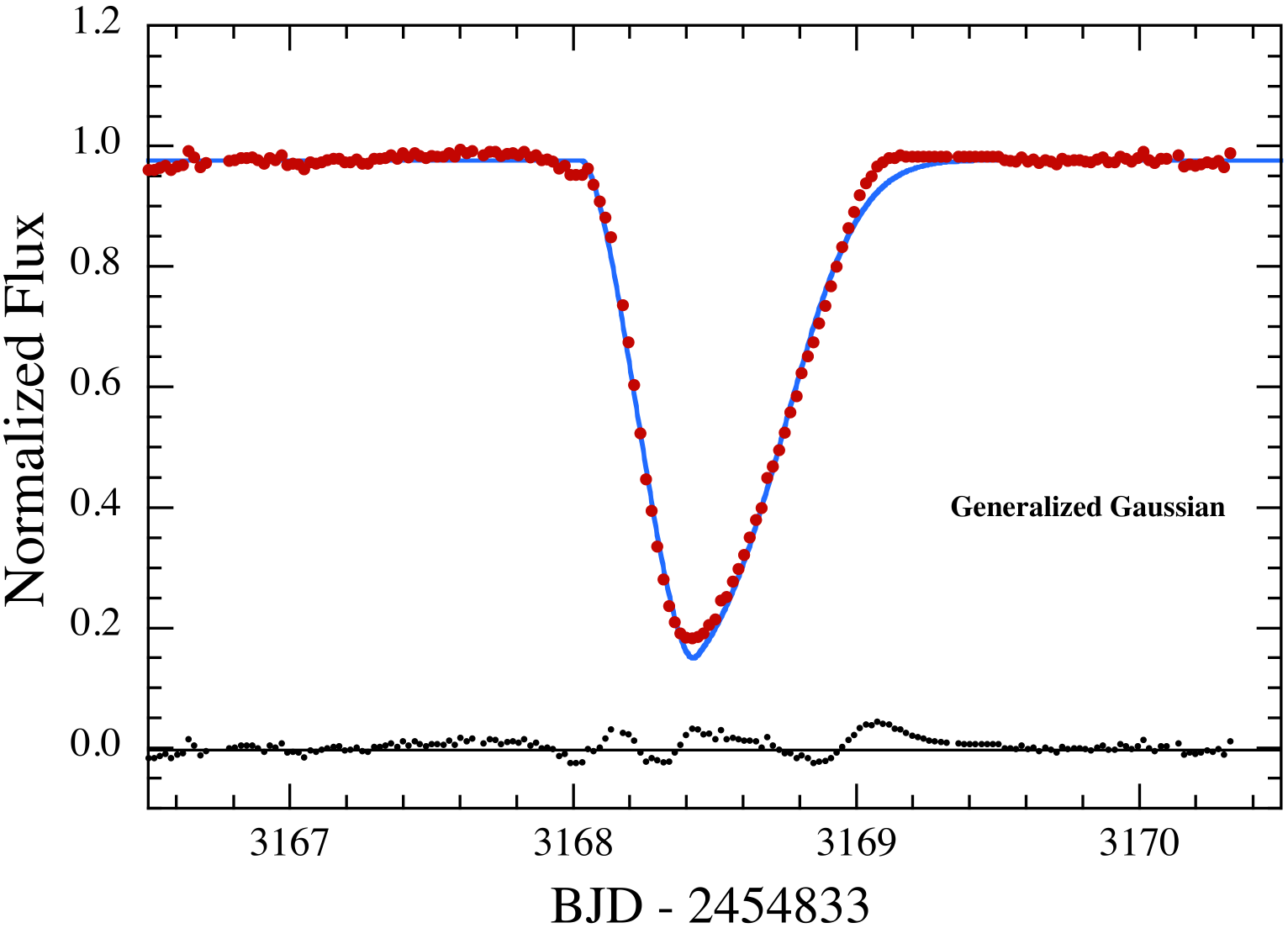}
\caption{Two model fits to the occultation.  Top panel: model consisting of a inclined and tilted disk of uniform optical depth (blue curve; see text) fit to the deep occultation of EPIC 204376071 (red data points).  Bottom panel: simple dust sheet model.  The optical depth behind the knife-edge is assumed to fall off with a generalized Gaussian profile (with $p \simeq 2.17$).  See text for details.  Each model has only six free parameters associated with it. The residuals from the best fitting model are shown at the bottom of each plot to the same scale as the data.}
\label{fig:modelfit}
\end{center}
\end{figure}

\begin{figure}
\begin{center}
\includegraphics[width=1.00 \columnwidth]{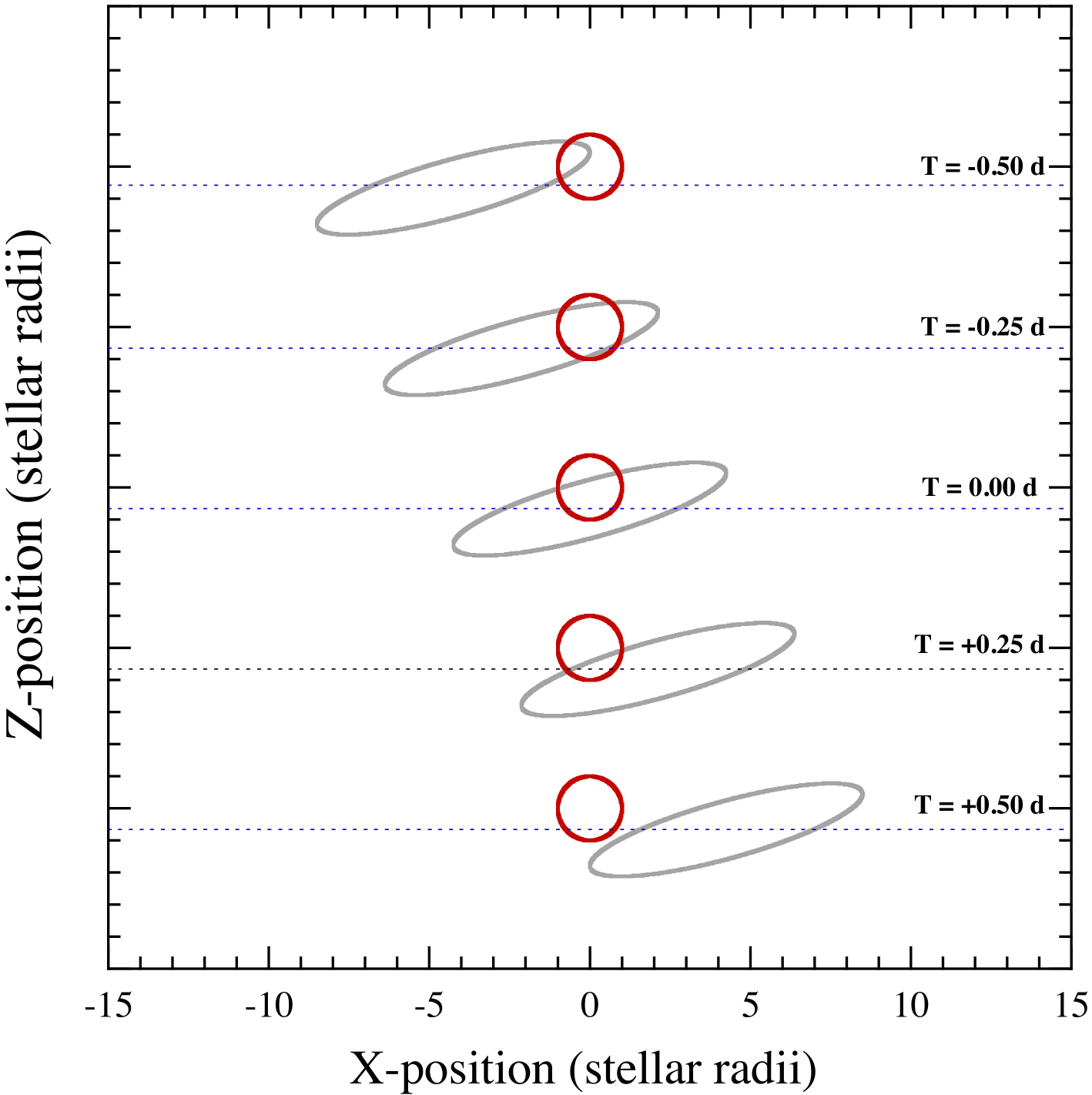}
\caption{Schematic, to scale, of a inclined and tilted disk, elliptical in projection, crossing its host star.  The snapshots at five different times illustrate how this geometry results in a distinctly asymmetric occultation.  Each snapshot is shifted vertically for clarity.  The dashed horizontal lines mark the trajectory of the disk center.}
\label{fig:scheme}
\end{center}
\end{figure}

The result of the fit to the disk occultation model is shown in the top panel of Fig.~\ref{fig:modelfit}, and the best-fit parameters are summarized with uncertainties in Table \ref{tbl:diskfit}.  The geometry of the occultation of the host star by the inclined and tilted disk is illustrated in Fig.~\ref{fig:scheme}.  It is relatively clear from the figure how such a disk can produce the desired highly asymmetric occultation.

In spite of the excellent quality of the fitted model, we can quickly see a potential physical problem with such a model.  The transverse speed during the occultation is 7.5 host star radii per day, which translates to 38 km s$^{-1}$.  For an assumed circular orbit this corresponds to an orbital period for the anchor body and disk of only $\sim$28 days\footnote{The lowest accepted value for $v$ in all of the MCMC chains was 6.2 $R_{\rm host}/d$ which translates to a circular orbit period of $\sim$50 days.} with an orbital radius of only 0.10 AU.  The problem, of course, is that orbital periods of $\lesssim 80$ days are ruled out.  Moreover, an anchor body of brown-dwarf proportions would be required so that the size of its Hill sphere for a circular orbit could accommodate such a large disk.  It is, of course, quite possible for the anchor body and disk to be in an eccentric orbit and have the occultation occur at periastron. The potential problems with the Hill sphere and too short orbital period are conveniently mitigated in the case of eccentric orbits.  See Sects.~\ref{sec:orbits} and \ref{sec:hills} for a detailed discussion of the eccentricities and periods required for the anchor-body companion to accommodate such a disk within its Hill sphere. 

\begin{table}
\centering
\caption{Fitted Parameters for the Inclined Disk Model$^a$}
\begin{tabular}{lc}
\hline
semimajor axis$^b$ [$R_{\rm host}$] & $4.2 \pm 0.3$ \\
semimajor axis$^b$ [AU] & $0.013 \pm 0.0009$ \\
semiminor axis$^b$ [$R_{\rm host}$] & $0.89 \pm 0.10$ \\
tilt angle$^c$ [deg] & $18.1 \pm 1.5$ \\
disk obliquity$^d$ [deg] & $22 \pm 2$\\
impact parameter, $b$  [$R_{\rm host}$] & $-0.88 \pm 0.07$ \\
$v_t$ [$R_{\rm host}/{\rm d}$] & $7.5 \pm 0.4$  \\
$v_t$ [km s$^{-1}$] & $38.3 \pm 3.4$  \\
$\tau$$^e$ & $2.5 \pm 0.4$  \\
$t_0$ [KBJD]$^f$ & $3168.72 \pm 0.02$  \\
DC$^g$ & $0.977 \pm 0.002$  \\ 
$\chi^2_\nu$$^h$ & $\equiv 1$  \\  
$P_{\rm orb}$ [d]$^i$ & $28 \pm 5$  \\
$a$ [AU]$^i$ & $0.10 \pm 0.01$  \\
\hline
\label{tbl:diskfit}
\end{tabular}

{\bf Notes.} (a) Inferred from an MCMC fit of the disk model to the dip (see Sect.~\ref{sec:specific}). (b) Apparent ellipse dimensions for the inclined circular disk. (c) The angle between the long axis of the ellipse and the direction of motion on the sky (see Fig.~\ref{fig:scheme}). (d) Tilt of the disk axis with respect to the anchor body's orbital angular momentum vector. This obliquity is comparable or less than for the rings of Saturn, Uranus, and Neptune. (e) Optical depth, assumed uniform over the disk.  (f) BKJD = BJD - 2454833. (g) The out of eclipse flux level. (h) The uncertainties are re-normalized so that $\chi^2$ is unity. (i) For an assumed circular Keplerian orbit.
\end{table}

\begin{table}
\centering
\caption{Fitted Parameters for the Dust-Sheet Model$^a$}
\begin{tabular}{lccc}
\hline
parameter & $\tau_0 e^{-x/\lambda}$ & $\tau_0 e^{-x^2/\lambda^2}$ & $\tau_0 e^{-x^p/\lambda^p}$ \\
\hline
power, $p$ & $\equiv 1$ & $\equiv 2$ & $2.17 \pm 0.20$ \\
$v_t$ [$R_{\rm host}/{\rm d}$] & $4.3 \pm 0.6$ & $4.6 \pm 0.4$ & $4.6 \pm 0.4$ \\
$v_t$ [km$^{-1}$] & $22.0 \pm 3.0$ & $23.5 \pm 2.0$ & $23.0 \pm 2.0$ \\
$\lambda$ [$R_{\rm host}$] & $1.33 \pm 0.2$ & $2.0 \pm 0.2$ & $2.0 \pm 0.2$ \\
$\tau_0$ & $2.5 \pm 0.3$ & $2.4 \pm 0.3$ & $2.4 \pm 0.3$ \\
$t_0$ [KBJD]$^b$ & $3168.23 \pm 0.01$ & $3168.23 \pm 0.02$ & $3168.23 \pm 0.02$ \\
DC$^c$ & $0.99 \pm 0.01$ & $0.98 \pm 0.01$ & $0.98 \pm 0.01$ \\ 
$\chi^2_\nu$$^d$ & 9.2 & 2.14 & $2.00$ \\ 
$P_{\rm orb}$ [d]$^e$ & $155 \pm 97$ & $123 \pm 50$ & $125 \pm 50$ \\
$a$ [AU]$^e$ & $0.31 \pm 0.11$ & $0.27 \pm 0.06$ & $0.27 \pm 0.06$ \\
\hline
\label{tbl:dustfit}
\end{tabular}

{\bf Notes.} (a) Inferred from an MCMC fit of the dust model to the dip (see Sect.~\ref{sec:specific}). The optical depth profile is given in the column heading. (b) BKJD = BJD - 2454833. (c) The out of eclipse level. (d) The $\chi^2$ values have been renormalized so that a value of unity is obtained for the disk model (see Table \ref{tbl:diskfit}). (e) For an assumed circular Keplerian orbit.
\end{table}

\subsection{Dust Sheet Model}
\label{sec:specific}

We next test a more generic occulter that involves a dust outflow made of small particles.  In this case, one would still require an orbiting body (e.g., a planet or brown dwarf) to either be, or provide the anchor for, the source of the dusty effluents.     

\begin{figure}
\begin{center}
\includegraphics[width=1.00 \columnwidth]{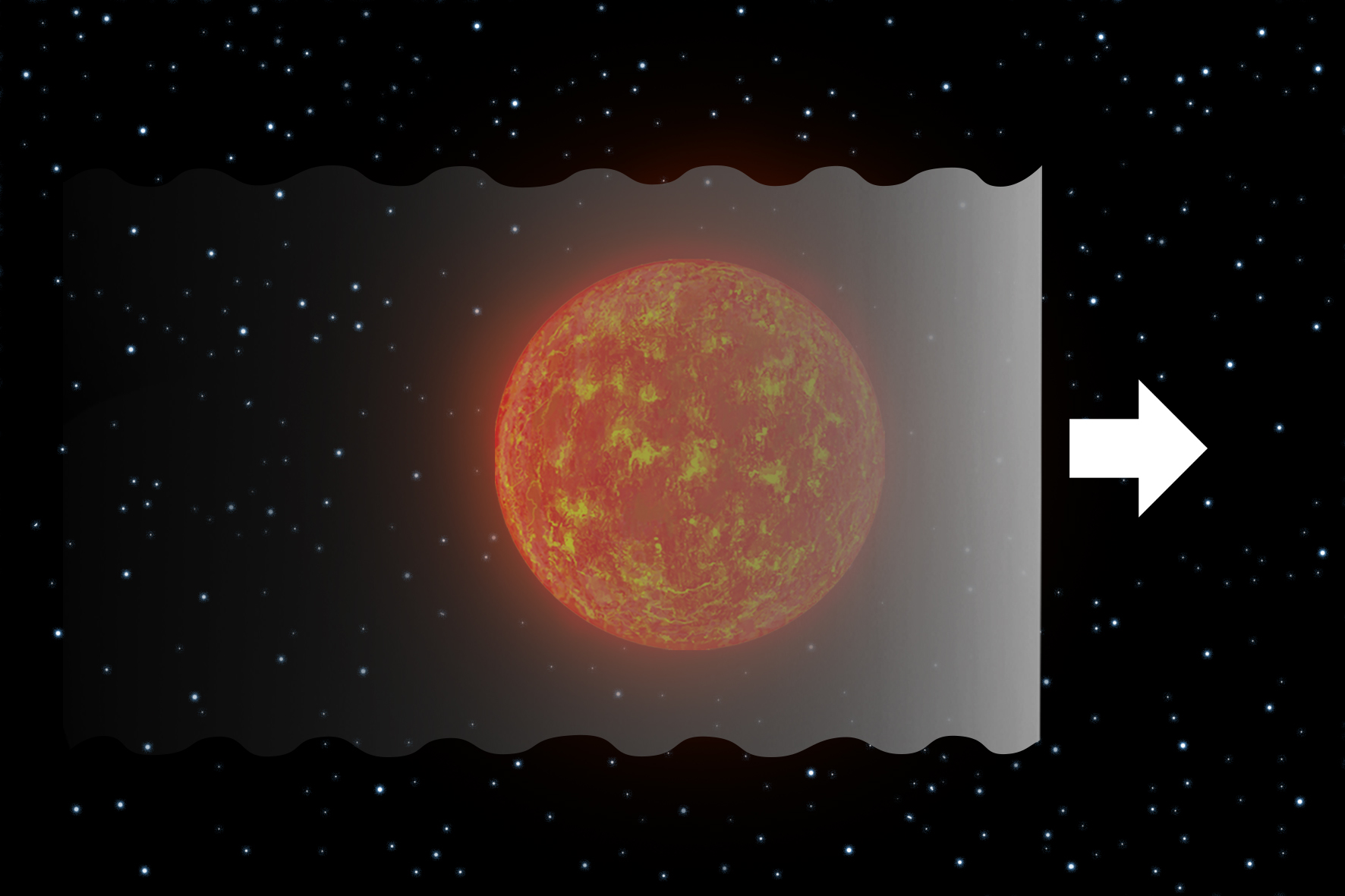}
\caption{Artist's conception of the knife-edge dust sheet passing in front of EPIC 204376071 (Credit: Danielle Futselaar; \url{http://www.artsource.nl/}.)}
\label{fig:artist}
\end{center}
\end{figure}

For the orbiting dust occulter we assume that the orbital period of the putative dust-trapping or dust-emitting body is substantially longer than the transit duration ($\sim$1 day), so that the transverse velocity, $v_t$, during the transit is essentially a constant.  For lack of more detailed information, we further assume that the dust ``sheet'' is wider than the size of the host star (otherwise, an 80\% occultation is difficult).  Finally, we assume that the optical depth  profile in the dust sheet is a simple function of distance from the leading edge of the dust sheet (see, e.g., \citealt{brogi12}; \citealt{sanchis-ojeda15} for a related geometry).  The geometry we are considering here is illustrated in the schematic of Fig.~\ref{fig:artist}. Of all the possibilities, we have experimented with only single-sided exponential, Gaussian, and generalized Gaussian profiles.

We take the location of the leading edge of the dust sheet, projected on the plane of the sky, to be at $x_o$, and an arbitrary location on the disk of the star to be $\{x,y\}$, where $\{0,0\}$ is the center of the stellar disk, and $x$ is the direction of motion of the dust sheet. We model the transmission, $T$, of starlight at $x$, as
\begin{eqnarray}
T(x,x_o) &=& \exp\left[- \tau_0 e^{-(x_o-x)/\lambda}\right] ~~x < x_o ~({\rm Expon.})  \nonumber \\
T(x,x_o)  & = &  \exp\left[- \tau_0 e^{-(x_o-x)^2/\lambda^2}\right] ~~x < x_o ~({\rm Gaussian}) \nonumber \\
T(x,x_o)  & = &  \exp\left[- \tau_0 e^{-(x_o-x)^p/\lambda^p}\right] ~~x < x_o ~({\rm G.G.}) \nonumber \\
\label{eqn:Gauss}
T(x,x_o)  & = & 0 ~~{\rm for}~x > x_o 
\end{eqnarray}
where $\tau_0$ is the optical depth at the leading edge of the dust sheet and $\lambda$ is the scale length for the falloff of the optical depth, $\tau$, within the dust sheet.  The labels ``Expon.'', ``Gaussian'', and ``G.G.'', stand for falloffs in the optical depths that are described by single-sided exponential, Gaussian, and generalized Gaussian functions, respectively. The position of the leading edge of the dust sheet along the $x$ direction is given as a function of time, $t$, by:
\begin{equation}
x_o = v_d (t-t_0)
\end{equation}
where $t_0$ is the time when the leading edge of the dust sheet crosses the center of the stellar disk, and $v_d$ is the speed of the dust sheet across the line of sight. We also adopt a quadratic limb darkening law for the host star, with coefficients appropriate to a mid-M star \citep{claret11}.  

We utilize the same Markov Chain Monte Carlo (`MCMC') code discussed above to fit a dust-sheet model to the occultation event.  In this case, there are five or six free parameters to be fit: $\tau_0$, $\lambda$, $v_t$, $t_0$, DC, and possibly $p$ for the case of the generalized Gaussian, where the ``DC'' term is again the background flux level away from the occultation. For each choice of parameters, we generate a model lightcurve by integrating over the stellar disk, including where the dust sheet overlaps the disk, and repeating this in increments of 6 minutes as the dust sheet crosses the stellar disk. The model lightcurve is then convolved with the $\sim$30 minute integration time of the {\em Kepler} long-cadence sampling. Here, each MCMC chain has $10^5$ links, and we have run a half-dozen chains.

The first conclusion we reached is that the exponentially distributed optical depth model does not produce an occultation profile that comes even close to matching the observed profile.  In fact, it produces a distinct shark-tooth shaped profile such as those that can sometimes be found in dipper sources \citep{ansdell16} and were used to fit transits of candidate exocomets \citep{rappaport18}. By contrast, the Gaussian optical depth profile gives a respectable, though imperfect, fit to the occultation.  The generalized Gaussian profile for the optical depth provides the best fit to the occultation (Fig.~\ref{fig:modelfit}; Table \ref{tbl:dustfit}).  The best-fitting power is $p \simeq 2.17$. The main shortcoming of this fit is that the model egress ends a bit more gradually than the profile seen in the $K2$ data (Fig.~\ref{fig:modelfit}).  Nonetheless, for a simple 6-parameter model, it seems to account for much of the occultation profile.  In principle, it is straightforward to adjust the optical depth profile with further parameters so as to match the data more perfectly; however, that seems unwarranted at this early stage of investigation.

The best-fitting parameters and their uncertainties for the exponential, Gaussian, and generalized Gaussian optical-depth profiles are summarized in Table \ref{tbl:dustfit}. In spite of the different quality of the fits for the different optical-depth profiles, the interesting physical parameters, such as $v_t$, $\lambda$, and $\tau_0$, do not vary very much.  In the following section we utilize the parameters derived from the generalized Gaussian model.

\section{Constraints on the System}
\label{sec:interpret}

The fits we have done for both an elliptically projected disk occulter and a dust sheet each yield a value for the transverse orbital speed during the occultation (see Tables \ref{tbl:diskfit} and \ref{tbl:dustfit}), while the {\em K2} photometry requires that the orbital period be $\gtrsim 80$ days.  As we will discuss here, there is also a constraint on the distance of closest approach to the host star that the dust can make before it is tidally shorn from its anchor body.  The goal here will be to use these values and constraints to determine what system parameters are allowed for the anchor body and host star.   

\subsection{Allowed Orbital Parameters}
\label{sec:orbits}

The most challenging parameter to accommodate is the relatively high transverse speed during the occultation for the disk model.  For a circular orbit, that speed ($\sim$38 km s$^{-1}$) leads to a prohibitively short orbital period.  Thus, we would need to invoke an eccentric orbit.  The least demanding orientation is then one where we observe the occultation when the disk anchor body is at periastron passage (with its largest orbital speed).  Thus, from hereon we work under the assumption that we are viewing the orbit with an argument of periastron equal to 270$^\circ$, i.e. with the transit taking place at periastron.

We start by writing down expressions for the periastron distance between the host star and anchor body and the corresponding orbital period:
\begin{eqnarray}
d_{\rm per} & = & \frac{G M_{\rm host} (1+e)}{v_t^2}   \nonumber \\ 
d_{\rm per} & \simeq & 34.9 \left(\frac{30 \,{\rm km/s}}{v_t}\right)^2 \,(1+e) ~R_\odot
\label{eqn:dmin}
\end{eqnarray}
where $M_{\rm host}$ is the mass of the host star, and is presumed to be much more massive than the body anchoring or producing the dust or disk, and $e$ is the orbital eccentricity.  The second line of this equation has been evaluated for the known mass of the host star and normalized to a transverse orbital speed of 30 km s$^{-1}$.  The corresponding orbital period is
 \begin{eqnarray}
P & = & \frac{2 \pi G M_{\rm host}}{v_t^3} \left(\frac{1+e}{1-e}\right)^{3/2} \nonumber \\
P & \simeq & 59.3 \left(\frac{30 \,{\rm km/s}}{v_t}\right)^3 \left(\frac{1+e}{1-e} \right)^{3/2} ~{\rm day} 
\label{eqn:period}
\end{eqnarray}
where the normalizations are the same as for Eqn.~(\ref{eqn:dmin}).  

For the model fit to a dust sheet, Table \ref{tbl:dustfit} indicates that $v_t \simeq 23$ km s$^{-1}$ in the best-fitting model.  Applying that value to Eqn.~\ref{eqn:period} yields $P \gtrsim 130$ d for any orbital eccentricity.  In turn, Eqn.~(\ref{eqn:dmin}) indicates that $d_{\rm per} \gtrsim 59 \, R_\odot = 0.28 $ AU. As we discuss below, this is sufficient for the instantaneous Hill sphere radius of a super-Earth to accommodate the hypothesized dust sheet. 

In the case of the elliptically projected disk occulter (see Fig.~\ref{fig:scheme}) the situation is not as simple.  Application of the best fitting transverse velocity of 38 km s$^{-1}$ to Eqn.~(\ref{eqn:period}) yields  $P \simeq 28.5 ((1+e)/(1-e))^{3/2}$ d, which requires a substantial eccentricity.  In particular, for $P$ to exceed the minimum value of 80 days, $e \gtrsim 0.33$. In turn, plugging this eccentricity into Eqn.~(\ref{eqn:dmin}) yields $d_{\rm per} \simeq 28.5 \,R_\odot \simeq 0.13$ AU. 

\subsection{Hill Sphere Radius of Anchor or Shepherding Body}
\label{sec:hills}

The size of the occulting disk or dust sheet must fit well within the instantaneous Hill sphere radius of the anchor or shepherding body at periastron passage:
\begin{equation}
R_h = \left(\frac{M_{\rm anch}}{3 M_{\rm host}} \right)^{1/3} d_{\rm per} ~,
\label{eqn:hill1}
\end{equation} 
where $M_{\rm anch}$ is the mass of the anchor or dust-shepherding body.  This can be rewritten in terms of $d_{\rm per}$ given by Eqn.~(\ref{eqn:dmin}): 
\begin{equation}
R_h = 34.9 \left(\frac{M_{\rm anch}}{3 \,M_{\rm host}} \right)^{1/3}  \left(\frac{30 \,{\rm km/s}}{v_t}\right)^2 \,(1+e)~R_\odot ~.
\label{eqn:hill2}
\end{equation}

For the case of the dust sheet, Eqn.~(\ref{eqn:hill2}) reduces to 
$$R_{\rm h} ~({\rm dust~sheet}) \simeq 2.4 \rightarrow 5.2 \,R_\odot \simeq 4 \rightarrow 8 \, R_{\rm host}$$
for $10 \, M_\oplus < M_{\rm anch} < 100\, M_\oplus$.  As mentioned above, this is just adequate to accommodate the dust sheet whose length is several times $R_{\rm host}$ (see Table \ref{tbl:dustfit}).

By contrast, in the case of the occulting disk, Eqn.~(\ref{eqn:hill2}), with $e = 0.33$, yields 
$$R_{\rm h} ~({\rm disk}) \simeq 1.2 \rightarrow 2.5 \,R_\odot \simeq 2 \rightarrow 4 \, R_{\rm host}$$
for the same range of $M_{\rm anch}$ as above.  However, since the best-fitting disk has a radius of 4.2 $R_{\rm host}$, and the Hill sphere radius should be at least twice this for stability, this requires a much greater anchor mass of $\sim$$3\,M_J$.  Thus, a super-Jupiter could do the job.  Again, this is only for the case where the orbital eccentricity is $\gtrsim 0.33$.

\subsection{Inferred Dust Mass and Mass Loss Rates} 
\label{sec:dust}                          

We now proceed to make a simple estimate of the mass required in small particles (dust to pebble size) that is required to obscure up to $\sim$80\% of the host star's light for a whole day.  We define three quantities from which everything else follows: the mass column density of the dust sheet, $\Sigma$, the bulk density of the small particles, $\rho$, and the characteristic size of the particles, $a$.  The column {\em number} density, mass, and cross section of the particles are: $N = \Sigma/\mu$, $\mu = 4\pi \rho a^3/3$,  and $\sigma = \xi \pi a^2$, respectively, where $\xi$ is the cross section normalized to the geometric area of a particle.

The optical depth of the dust is
\begin{equation}
\tau = N\sigma = \frac{1}{\mu} \xi \pi a^2 \Sigma = \frac{3 \xi}{4} \frac{\Sigma}{\rho a}  ~.
\end{equation}

For a uniform density dust disk the mass is just $A_{\rm disk} \Sigma$:
 \begin{equation}
\mathcal{M}_d \simeq  4 \rho \,a \,\tau A_{\rm disk}/(3 \xi)  ~.
\label{eqn:mass1}
\end{equation}
where $A_{\rm disk}$ is the projected area of the inclined disk (i.e., seen as an ellipse).
If we then adopt the following nominal set of parameter values: $A_{\rm disk} \simeq \pi \, (4.2 \times 0.89) \, R_{\rm host}^2$, $\tau \simeq 2.9$ (see Table \ref{tbl:diskfit}), $\rho = 3$ g cm$^{-3}$, $a = 1\, \mu$m, and $\xi = 1$, we find a required minimum mass for such a disk to be:
$$ \mathcal{M}_d \simeq 2.6 \times 10^{19}~{\rm g}~~.$$ 
Here we have assumed $\mu$m-sized particles for the obscuring material since that leads to a minimum required mass.  For smaller particles, $\xi$ can become dramatically smaller than unity, while for larger particles, $a$ grows while $\xi \rightarrow $$\sim$1.  However, in either case, we can see from Eqn.~(\ref{eqn:mass1}) that the required mass will rise.

We note in passing that the mass in Saturn's rings is estimated to be $\sim$1000 times the value we infer for $\mathcal{M}_d$.  Moreover, Saturn's rings would barely be large enough to block 10\% of the light of EPIC 204376071 even if they were all optically thick.  However, much of the mass in Saturn's rings are in particles that are $\gg 1 \,\mu$m (see, e.g., \citealt{zebker85}), and therefore very inefficient, per unit mass, of blocking light.

For the case of the dust sheet, we base the mass on the generalized Gaussian profile with optical depth profile given by Eqn.~(\ref{eqn:Gauss}).  The optical depth at any point within the dust sheet is
\begin{equation}
 \tau  =  \tau_0 \,e^{-X^p/\lambda^p} = \frac{3 \xi}{4} \frac{\Sigma_0}{\rho a} e^{-X^p/\lambda^p}  ~,
\end{equation}
where $\Sigma_0$ is the mass column density right behind the leading edge of the dust sheet and $X$ is an arbitrary distance behind the edge.  

The total mass contained in the dust sheet is given by
\begin{eqnarray}
\mathcal{M}_d & > & \int_0^\infty 2 R_{\rm host} \, \Sigma_0 \, e^{-X^p/\lambda^p} dX  \nonumber \\
& \simeq & \frac{8\, \Gamma[(1+p)/p)]}{3 \xi} R_{\rm host} \,\rho \, a \, \tau_0 \, \lambda ~,
\label{eqn:mass2}
\end{eqnarray}
where the parameters $\tau_0$, $\lambda$, and $p$ emerge directly from the fitting process (see Table \ref{tbl:parms}), and $\Gamma$ is a gamma-function.  The inequality takes into account the fact that the putative dust sheet may extend much beyond a width of $2 R_{\rm host}$.  If we take the following nominal set of parameter values: $R_{\rm host} \simeq 0.63 \, R_\odot$, $\rho = 3$ g cm$^{-3}$, $a = 1\, \mu$m, $\tau_0 \simeq 2.5$, $\lambda \simeq 2.0 \, R_\odot$, $p = 2.17$ (see Table \ref{tbl:dustfit}), and $\xi = 1$, then we find a required minimum mass for an orbiting dust sheet to be:
$$ \mathcal{M}_d \simeq 1 \times 10^{19}~{\rm g}~~.$$ 

This mass is a factor of $\sim$3 lower than required for the dusty disk, largely because the disk was found to be several times larger in linear dimension that the dust sheet.

\subsection{Possible `Dipper' Scenario}
\label{sec:dipper}

Because EPIC 204376071 is a member of the Upper Sco association, and many of the so-called `dipper stars' are found there (see, e.g., \citealt{cody14}; \citealt{ansdell16}), the possibility arises as to whether the large occultation event we have found is also related to the `dipper' phenomena. In that case, the occultation would probably occur near the star and be due to some dusty material orbiting near the corotation radius of the host star.  While we cannot rule out that possibility, there are three properties of EPIC 204376071 which make it quite different from the other `dippers'.  

First, if we examine the numerous dipper stars presented in \citet{cody14} and \citet{ansdell16} we see that the dips in flux in those stars are typically 10\% to 30\% deep, with a few reaching depths of 50\%-60\%.  However, we note that the dipper-like behavior of HQ Tau \citep{rodriguez17b} has dips that equal or exceed the depth of the occultation in EPIC 204376071, and their durations are two to three orders of magnitude longer than for EPIC 204376071.  Second, almost all the dippers show some activity during much of their lightcurves which typically span 40-80 days in duration with satellite observations, or up to a decade with ground-based photometry.  None of them exhibits a single dip in flux, though that may be a selection effect, i.e., the reason they might not have been chosen for inclusion as a dipper.  Finally, many of the dipper host stars exhibit clear excess emission in the WISE 3 and 4 bands (see, e.g., \citealt{ansdell16}).  EPIC 204376071 has weak WISE 3 band emission and only an upper limit in the WISE 4 band (see Sect.~\ref{sec:spectra}).

\section{Summary and Conclusions}
\label{sec:summary}

In this work we report the discovery of a deep depression in the flux (by 80\%) of EPIC 204376071 that lasts for a full day.  The lightcurve of EPIC 204376071 is otherwise quiet for a total of 160 days of observation during the $K2$ campaigns C2 and C15.  

The host star is a low-mass M-star ($\sim$0.16\,M$_\odot$) with a radius of $\sim$0.63\,R$_\odot$ that is almost certainly in the Upper Scorpius stellar association with an age of $\sim$10 Myr.  The large radius compared to its nominal main-sequence radius indicates that the star has not yet settled onto the main sequence.  The rotation period of 1.6 days is consistent with the star's youth.

We have explored two basic scenarios for producing a deep asymmetric occultation of the type observed in EPIC 204376071. In the first we considered an intrinsically circular disk of dusty material anchored to a minor body orbiting the host star. The disk is inclined to the observer so that, in projection, it appears as a highly elliptical occulter. If taken to be of uniform optical depth, for simplicity, we find that the disk has a diameter of $\simeq 4.2$ times the radius of the host star.  Such a disk contains $\gtrsim 3 \times 10^{19}$ g of material. The fit to the measured occultation profile is excellent.  The body that anchors such a disk would have to be in an eccentric orbit and have a mass of $\sim$$3\,M_J$. It is interesting to note that the other occultation event for which a disk yields a good fit (1SWASP J140747 discussed in the Introduction; see~\citealt{kenworthy15a}) is also in the Scorpius-Centaurus Association.  The disk models for J1407 had to contend with the same issue, namely that the best fit appears to necessitate eclipses near periastron passage in a substantially eccentric orbit.

Second, we considered a dust sheet of material of basically unknown origin, though we do assume that the source of the dust is in a quasi-permanent orbit about the host star. In such a model we take the optical depth of the dust sheet to vary systematically with distance behind its leading edge (see Fig.~\ref{fig:artist} for an artist's conception). The fit to the occultation profile using this simple model also yields a good fit, but with a small systematic deviation during part of the egress.  Such a dust sheet requires somewhat less material at $\sim$$10^{19}$ g of material.

The possibility of a dust tail streaming from a minor body (e.g., such as KIC 1255b \citealt{rappaport12}) is hard to accept in the case of EPIC 204376071 given that there would be so little radiant flux at distances of $\gtrsim 1/4$ AU from a host star with $L \lesssim 0.027$\,L$_\odot$ (see \citealt{vanlieshout18}, for a discussion of dust production from small planets).  The equilibrium temperature, $T_{\rm eq}$, in such an environment would be $\approx 230$ K, which would be sufficiently low that the region would be closer to habitable than being prone to producing dusty effluents.

Any possible period to the dips, if there is one, is quite long (i.e., $\gtrsim 100$ days), and since the star is relatively faint, unfortunately the prospects for long-term ground-based monitoring of this K$_p = 15$ and V = 16.3 magnitude star are not great.  Therefore, it will be quite difficult to prove one way or the other whether the dip observed in EPIC 204376071 is part of a periodic set of dips, or an isolated event. 

Some possible shortcomings of a quasi-permanent orbiting dusty feature (see Sect.~\ref{sec:interpret}) have led us to also consider the possibility that this dip is `simply' an accretion event containing a large amount of dusty material.  After all, the target star is part of a stellar association replete with many other `dipper' stars.  A range of dipper-like stars and related objects was discussed in the Introduction (Sect.~\ref{sec:intro}).  It is true that the lightcurve of EPIC 204376071 is fairly unique among the group of dippers in that it is quiet for so long, and is only punctuated by a well-defined large drop in flux for just one day.  Nonetheless, we have to allow for the possibility that this dip is `merely' a large dusty accretion event.  It would certainly be more exciting if there is a quasi-permanent orbiting dusty structure in the system, but this remains to be proven.

Long-term ground-based monitoring for subsequent dips is not very likely in the medium term. The two {\em K2} campaigns observed dips in EPIC~204376071 with a probability of $\sim$1 in 160. With $\sim$1000 visits per field in the nominal LSST survey \citep{ivezic08}, this large survey would expect to see another dip in its 10 years of operations, unless the {\em K2} campaigns were particularly lucky in catching the single dip.  In addition, TESS may observe EPIC 204376071 during an extended mission \citep{ricker14,bouma17}.  In any case, it would be possible to make some very helpful radial velocity measurements to search for evidence of an orbiting body, and adaptive optics observations to search for scattered-light from disk structures or evidence of low-mass wide companions.   Orbiting bodies of 1 $M_J$, 10 $M_J$, and 40 $M_J$ in a 100-day orbit would yield RVs amplitudes of 150 m s$^{-1}$, 1.5 km s$^{-1}$ and 5 km s$^{-1}$, respectively.  Such measurements would require a large telescope with a red-sensitive spectrograph for radial velocity observations given the host star's K$_p = 15$ magnitude, and laser guide star or infrared wavefront sensor measurements in the case of adaptive optics, but would be worth the investment if an orbiting body is found. 

\vspace{20pt}

\noindent
{\bf Acknowledgements} 

We are grateful to the referee, Hugh Osborn, for extremely helpful suggestions in generally improving the manuscript, and for pointing the way to a successful disk fit to the occultation.
Support for G.\,Z.~is provided by NASA through Hubble Fellowship grant HST-HF2-51402.001-A awarded by the Space Telescope Science Institute, which is operated by the Association of Universities for Research in Astronomy, Inc., for NASA, under contract NAS 5-26555. 
A.\,V.'s work was supported in part under a contract with the California Institute of Technology (Caltech)/Jet Propulsion Laboratory (JPL) funded by NASA through the Sagan Fellowship Program executed by the NASA Exoplanet Science Institute.
E.\,R.\,N.~is supported by an NSF Astronomy and Astrophysics Postdoctoral Fellowship under award AST-1602597
We thank Allan R. Schmitt and Troy Winarski for making their lightcurve examining software tools {\tt LcTools} and {\tt AKO-TPF} freely available.  
Some of the data presented in this paper were obtained from the Mikulski Archive for Space Telescopes (MAST). STScI is operated by the Association of Universities for Research in Astronomy, Inc., under NASA contract NAS5-26555. Support for MAST for non-HST data is provided by the NASA Office of Space Science via grant NNX09AF08G and by other grants and contracts. 
Some results are based on data from the Carlsberg Meridian Catalogue 15 Data Access Service at CAB (INTA-CSIC). 
This research has made use of IMCCE's SkyBoT VO tool

\end{document}